\documentclass[11pt,onecolumn,draftcls]{IEEEtran}
\usepackage{amsmath,amssymb}
\usepackage{graphicx}
\usepackage{verbatim}
\usepackage{multirow}
\usepackage{enumerate}
\usepackage{cite}

\title{Composite Cyclotomic Fourier Transforms \\with Reduced
  Complexities}

\author{Xuebin~Wu, Meghanad~Wagh, Ning~Chen, Ying~Wang, and
  Zhiyuan~Yan

  \thanks{This work was supported in part by NSF under grant
    ECCS-0925890 and in part by a grant from Thales Communications
    Inc. The material in this paper was presented in part at the 2010
    IEEE Workshop on Signal Processing Systems, Cupertino, CA, in
    October 2010.}

  \thanks{Xuebin Wu, Meghanad Wagh, and Zhiyuan Yan are with the
    department of ECE, Lehigh University, PA 18015,
    USA. E-mails:\{xuw207, mdw0, yan\}@lehigh.edu.}

  \thanks{Ning Chen was with the department of ECE, Lehigh University,
    and now is with PMC-Sierra, Allentown, PA 18104, E-mail:
    ning\_chen@pmc-sierra.com.}

    \thanks{Ying Wang is with Qualcomm Flarion Technologies,
      Bridgewater, NJ 08807, USA. E-mail: aywang11@gmail.com.}}

\renewcommand{\L}{\mathbf{L}}
\newcommand{\f}{\mathbf{f}}
\newcommand{\Q}{\mathbf{Q}}

\newcommand{\A}{\mathbf{A}}

\begin{document}
\maketitle
\begin{abstract}
  Discrete Fourier transforms~(DFTs) over finite fields have
  widespread applications in digital communication and storage
  systems.  Hence, reducing the computational complexities of DFTs is
  of great significance.
  Recently proposed cyclotomic fast Fourier
  transforms (CFFTs) are promising due to their low multiplicative
  complexities. Unfortunately, there are two issues with CFFTs: (1)
  they rely on efficient short cyclic convolution algorithms, which
  has not been sufficiently investigated in the literature,  and (2)
  they have very high additive complexities when directly implemented.
  To address both issues, we make three main contributions in this
  paper. First, for any odd prime $p$, we reformulate a $p$-point
  cyclic convolution as a product of a $(p-1)\times (p-1)$ Toeplitz
  matrix vector  products (TMVP), which can be obtained from well-known
  TMVP of very small sizes, leading to efficient bilinear algorithms
  for $p$-point cyclic convolutions.
  Second, to address the high additive complexities of CFFTs, we
  propose composite cyclotomic Fourier transforms (CCFTs).
  In comparison to previously proposed fast Fourier transforms, our
  CCFTs achieve lower overall complexities for moderate to long
  lengths, and the improvement significantly increases as the length
  grows. Third, our efficient algorithms for $p$-point cyclic
  convolution and CCFTs allow us to obtain longer DFTs over larger
  fields, e.g., 2047-point DFT over GF$(2^{11})$ and 4095-point DFT
  over GF$(2^{12})$, which are first efficient DFTs of such lengths to
  the best of our knowledge.
  Finally, our CCFTs are also advantageous for hardware
  implementations due to their regular and modular structure.
\end{abstract}


\section{Introduction}
Discrete Fourier transforms~(DFTs) over finite fields
\cite{BlahutFast} have widespread applications in error correction
coding, which in turn is used in all digital communication and storage
systems. For instance, both syndrome computation and Chien search in
the syndrome based decoder of Reed-Solomon codes
\cite{BlahutECC,Wicker95}, a family of widely used error control
codes, can be formulated as polynomial evaluations and hence can be
implemented efficiently using DFTs over finite fields. Implementing an
$N$-point DFT directly requires $O(N^2)$ multiplications and $O(N^2)$
additions, and becomes costly when $N$ is large. Hence, reducing the
computational complexities of DFTs is of great significance. Recently,
efficient long DFTs have become particularly important as increasingly
longer error control codes are chosen for digital communication and
storage systems. For example, Reed-Solomon codes over GF$(2^{12})$ and
with block length of several thousands are considered for hard drive
\cite{IDEMAWhitePaper} and tape storage \cite{Han2005} as well as
optical communication systems \cite{Buerner2004} to achieve better
error performance; the syndrome based decoder of such codes requires
DFTs of lengths up to $4095$ over GF$(2^{12})$. In addition to
complexity, regular and modular structure of DFTs is desirable for
efficient hardware implementations.

In the literature, fast Fourier transforms (FFTs) based on the
prime-factor algorithm \cite{Good1958} and the Cooley-Turkey algorithm
\cite{Cooley1965} have been proposed for DFTs over the complex
field. When FFTs based on the prime-factor algorithm are adapted to
DFTs over finite fields \cite{Truong2006}, they still have high
multiplicative complexities. In contrast, recently proposed cyclotomic
FFTs (CFFTs) are promising since they have significantly lower
multiplicative complexities \cite{Trifonov2003,Fedorenko2006}.
However, CFFTs have two issues. First, they rely on efficient
algorithms for short cyclic convolutions, which do not always
exist. For instance, CFFTs over GF$(2^{11})$ would require efficient
algorithms for $11$-point cyclic convolutions. Previous works (see,
for example, \cite{Trifonov2003,Fedorenko2006,Chen2009c}) have not
investigated CFFTs over GF$(2^{11})$ partially due to the lack of
efficient $11$-point cyclic convolutions in the literature. Second,
CFFTs have very high additive complexities when directly implemented,
which can be reduced by techniques such as the common subexpression
elimination (CSE) (see, for example,
\cite{Trifonov2007,Aho1974,Oscar2007,Chen2009c}). In particular, the
CSE algorithm in \cite{Chen2009c} is effective for reducing the
additive complexities of CFFTs over GF$(2^l)$ for $l\le 10$. However,
although the CSE algorithm has a polynomial complexity
\cite[Sec.~III-F]{Chen2009c}, its time and memory requirements limit
its effectiveness for long DFTs. Due to these two issues, CFFTs over
GF$(2^{11})$ and GF$(2^{12})$ have not been investigated in the
literature.

In this paper, we address both aforementioned issues. The main contributions of our paper are as follows.
\begin{itemize}
\item For an odd prime $p$, we reformulate a $p$-point cyclic
  convolution over characteristic-$2$ finite fields as a product of a
  $(p-1)\times (p-1)$ Toeplitz matrix and a vector. Since $p-1$ is
  composite, this product can be readily obtained by multi-dimensional
  technology from well-known Toeplitz matrix vector products (TMVP) of
  very small sizes
  \cite{Allwright1971,Pitassi1971,Agarwal1974,Winograd1980,Fan2007}. In
  comparison to other ad hoc techniques based on TMVP, our
  reformulation achieves lower multiplicative complexity, especially
  for small to moderate $p$. Hence, our reformulation leads to
  efficient bilinear algorithms for $p$-point cyclic convolution over
  characteristic-$2$ finite fields. Our reformulation can be readily
  extended to the real and complex fields as well as more general
  finite fields. Furthermore, by multi-dimensional technology, we can
  also obtain efficient algorithms for $p^n$-point cyclic
  convolutions. These algorithms are also key to long CFFTs.
\item Due to the high additive complexities of CCFTs, we propose
  composite cyclotomic Fourier transforms (CCFTs), which are
  generalization of CFFTs. When the length $N$ of the DFT is factored,
  that is, $N=N_1\times N_2$, our CCFTs use $N_1$- and $N_2$-point
  CFFTs as sub-DFTs via the prime-factor and Cooley-Turkey
  algorithms. Thus, CFFTs are simply a special case of our CCFTs,
  corresponding to the trivial factorization, i.e., $N=1\times N$.
This generalization reduces overall complexities in three ways. First,
this divide-and-conquer strategy itself leads to lower
complexities. Second, the moderate lengths of the sub-DFTs enable us
to apply complexity-reducing techniques such as the CSE algorithm in
\cite{Chen2009c} more effectively. Third, when the length $N$ admits
different factorizations, the one with the lowest complexity is
selected.  In the end, while an $N$-point CCFT may have a higher
multiplicative complexity than an $N$-point CFFT, the former achieves
a lower overall complexity for long DFTs because of its significantly
lower additive complexity. Moreover, when $N$ is composite, an
$N$-point CCFT has a regular and modular structure, which is suitable
for efficient hardware implementations.  Our CCFTs provide a systematic
  approach to designing long DFTs with low complexity.

\item Our efficient algorithms for $p$-point cyclic convolution and
  CCFTs allow us to obtain longer CFFTs over larger fields. For
  example, we propose CFFTs over GF$(2^{11})$, which are unavailable
  in the literature heretofore partially due to the lack of efficient
  $11$-point cyclic convolution algorithms.  Our 2047-point DFTs over
  GF$(2^{11})$ and 4095-point DFTs over GF$(2^{12})$ are also first
  efficient DFTs of such lengths to the best of our knowledge, and
  they are promising for emerging communication systems.
\end{itemize}

Our work in this paper extends and improves previous works
\cite{Trifonov2003,Chen2009c} on CFFTs over finite fields of
characteristic two in several ways.  First, previously proposed CFFTs
focus on $(2^l-1)$-point CFFTs over GF$(2^l)$ for $l\leq 10$. In
contrast, our CCFTs allow us to derive long DFTs with low complexity
over larger fields. Our approach can be applied to any finite field,
but we present CCFTs over GF$(2^{11})$ and GF$(2^{12})$ due to their
significance in applications. Furthermore, our work investigates
$N$-point CFFTs over GF$(2^l)$ for $N|2^{l}-1$.  Second, our CCFTs
achieve lower overall complexities than \textbf{all} previously
proposed FFTs for moderate to long lengths, and the improvement
significantly increases as the length grows.

The rest of the paper is organized as
follows. Sec.~\ref{sec:background} briefly reviews the necessary
background of this paper, such as the CFFT, the prime-factor
algorithm, the Cooley-Turkey algorithm, and the CSE algorithm. We
propose an efficient bilinear algorithm for $p$-point cyclic
convolutions over GF$(2^l)$ in Sec.~\ref{sec:conv}. We then use an
$11$-point cyclic convolution algorithm to construct 2047-point CFFT
over GF$(2^{11})$ in Sec.~\ref{sec:CCFT}. We also propose our CCFTs
and compare their complexities with previously proposed FFTs in
Sec.~\ref{sec:CCFT}. The advantages of our CCFTs in hardware
implementations are discussed in Sec.~\ref{sec:arch}. Concluding
remarks are provided in Sec.~\ref{sec:conclusion}.

\section{Background}
\label{sec:background}

\subsection{Cyclotomic Fast Fourier Transforms}
In this paper, we consider DFTs over finite fields of characteristic
two. Let $\alpha \in \mbox{GF}(2^{l})$ be an element with order $N$, which
implies that $N|2^{l}-1$ (otherwise $\alpha$ does not exist). Given an
$N$-dimensional column vector $\mathbf{f}=(f_0,f_1, \cdots,
f_{N-1})^T$ over GF$(2^l)$, the DFT of $\mathbf{f}$ is given by
$\mathbf{F}=(F_0, F_1, \cdots, F_{N-1})^T$, where
\begin{equation}
  \label{eq:DFT}
  F_j=\sum_{i=0}^{N-1}f_i\alpha^{ij}.
\end{equation}
If we define $f(x)=\sum_{i=0}^{N-1}f_ix^i$, we have $F_j=f(\alpha^j)$.
Directly computing the DFT requires $O(N^2)$ multiplications and
$O(N^2)$ additions, and is impractical for large $N$s. Cyclotomic FFTs
(CFFTs) \cite{Trifonov2003,Fedorenko2006} can reduce the
multiplicative complexities greatly.

We first partition the integer set $\{0, 1, \cdots, N-1\}$ into $m$
cyclotomic cosets modulo $N$ with respect to GF$(2)$ \cite{Wicker95}:
$C_{s_0}, C_{s_1}, \cdots, C_{s_{m-1}}$, where $C_{s_k}= \{2^0s_k,
2^1s_k, \cdots, 2^{m_k-1}s_k\} \pmod{N}$ and
$s_k=2^{m_k}s_k\pmod{N}$. A polynomial $L(x)=\sum_il_ix^{2^i}$, where
$l_i \in \mathrm{GF}(2^l)$, is called a \emph{linearized polynomial}
over GF$(2^l)$, since it has a linear property $L(x+y)=L(x)+L(y)$ for
$x, y \in \mathrm{GF}(2^l)$. With the help of cyclotomic cosets,
$f(x)$ can be decomposed as a sum of linearized polynomials
$$f(x)=\sum_{k=0}^{m-1}L_k(x^{s_k}),\quad
L_k(x)=\sum_{j=0}^{m_k-1}f_{s_k2^j\mathrm{mod}\,N}x^{2^j}.$$ Therefore
$F_j=\sum_{k=0}^{m-1}L_k(\alpha^{js_k})$, and each $\alpha^{js_k}$
lies in the subfield GF$(2^{m_k}) \subseteq $ GF$(2^{l})$.

Using a normal basis $\{\gamma_k^{2^0}, \gamma_k^{2^1}, \cdots,
\gamma_k^{2^{m_k-1}}\}$ in GF$(2^{m_k})$, $\alpha^{js_k}$ can be
expressed by $\sum_{i=0}^{m_k-1}a_{i,j,k}\gamma_k^{2^i}$, where
$a_{i,j,k}\in\{0,1\}$. By the linear property of $L_i(x)$'s, $ F_j
=\sum_{k=0}^{m-1}\sum_{i=0}^{m_k-1}a_{i,j,k}L_k(\gamma_k^{2^i})$. Written
in the matrix form, the DFT of $\mathbf{f}$ is given by $\mathbf{F}=
\mathbf{A}\mathbf{L}\mathbf{\Pi}\mathbf{f}$, where $\mathbf{A}$ is an
$N\times N$ binary matrix constructed from the binary coefficients
$a_{i,j,k}$, $\mathbf{\Pi}$ is an $N\times N$ permutation matrix,
$\mathbf{L}=\mathrm{diag}(\L_0, \L_1, \cdots, \L_{m-1})$ is a block
diagonal matrix, and $\mathbf{L}_k$'s are $m_k\times m_k$ square
matrices. The permutation matrix $\mathbf{\Pi}$ reorders the vector
$\mathbf{f}$ into $\mathbf{f}'=(\mathbf{f'}_0^T, \mathbf{f'}_1^T,
\cdots, \mathbf{f'}_{m-1}^T)^T$, and
$\mathbf{f'}_k=(f_{s_k2^0\mathrm{mod} N}, f_{s_k2^1 \mathrm{mod} N},
\cdots, f_{s_k2^{m_k-1} \mathrm{mod} N})^T$.

Though the idea of cyclotomic decomposition dates back to
\cite{Zakharova1992}, the normal basis representation is a key step
\cite{Trifonov2003}. Since $\gamma_k^{2^{m_k}}=\gamma_k$, the $k$-th
block $\mathbf{L}_k$ of $\mathbf{L}$ is actually a circulant matrix,
which is given by
\begin{equation*}
\L_k=\left[
  \begin{array}{cccc}
    \gamma_k^{2^0}      & \gamma_k^{2^1} & \cdots & \gamma_k^{2^{m_k-1}}\\
    \gamma_k^{2^1}      & \gamma_k^{2^2} & \cdots & \gamma_k^{2^{0}}\\
    \vdots             & \vdots        & \ddots & \vdots         \\
    \gamma_k^{2^{m_k-1}} & \gamma_k^{2^0} & \cdots & \gamma_k^{2^{m_k-2}}
  \end{array}
\right].
\end{equation*}
Hence the multiplication between $\L_k$ and $\f_k'$ can be formulated
as an $m_k$-point cyclic convolution between
$\mathbf{b}_k=(\gamma_k^{2^0}, \gamma_k^{2^{m_k}-1},
\gamma_k^{2^{m_k}-2} \cdots, \gamma_k^{2^{1}})^T$ and $\f_k'$. Since
$m_k$ is usually small, we can use efficient bilinear form algorithms
\cite{BlahutFast} for short cyclic convolutions to compute
$\mathbf{L}_k\mathbf{f'}_k$. Those bilinear form algorithms have the
following form,
$$\L_k\f_k'=\mathbf{b}_k\otimes\f_k'=\mathbf{Q}_k(\mathbf{R}_k\mathbf{b}_k\cdot\mathbf{P}_k\f_k')
= \mathbf{Q}_k(\mathbf{c}_k\cdot\mathbf{P}_k\f_k'),$$ where
$\mathbf{P}_k$, $\mathbf{Q}_k$, and $\mathbf{R}_k$ are all binary
matrices, $\mathbf{c}_k=\mathbf{R}_k\mathbf{b}_k$ is a precomputed
constant vector, and $\cdot$ denotes an component-wise multiplication
between two vectors. Combining all the matrices, we get
\begin{equation}
  \label{eq:CFFT}
  \mathbf{F}=\mathbf{A}\mathbf{Q}(\mathbf{c}\cdot\mathbf{P}\f'),
\end{equation}
where $\Q=\mathrm{diag}(\Q_0,\Q_1, \cdots, \Q_{m-1})$,
$\mathbf{P}=\mathrm{diag}(\mathbf{P}_0, \mathbf{P}_1,$ $\cdots,
\mathbf{P}_{m-1})$, and
$\mathbf{c}=(\mathbf{c}_0^T,\mathbf{c}_1^T,\cdots,
\mathbf{c}_{m-1}^T)^T$.

The multiplications required by \eqref{eq:CFFT} are due to the
component-wise multiplication between $\mathbf{c}$ and $\mathbf{Pf}'$,
and the additions required by \eqref{eq:CFFT} are for multiplications
between binary matrices and vectors.  Direct implementation of CFFT in
\eqref{eq:CFFT} requires much fewer multiplications than the direct
implementation of DFT, at the expense of a very high additive
complexity.


\subsection{Common Subexpression Elimination}
Given an $N\times M$ binary matrix
$\mathbf{M}$ and an $M$-dimensional vector $\mathbf{x}$ over a field
$\mathbb{F}$. The matrix vector  multiplication $\mathbf{Mx}$ can be
done by additions over $\mathbb{F}$ only, the number of which is
denoted by $\mathcal{C}(\mathbf{M})$ since the complexity is
determined by $\mathbf{M}$, when $\mathbf{x}$ is arbitrary. The problem of determining the minimal
number of additions, denoted by
$\mathcal{C}_\mathrm{opt}(\mathbf{M})$, has been shown to be
NP-complete \cite{Carey1979}.
Instead, different common subexpression elimination algorithms (see,
e.g., \cite{Trifonov2007,Aho1974,Oscar2007}) have been proposed to
reduce $\mathcal{C}(\mathbf{M})$.  The CSE algorithm proposed in
\cite{Chen2009c} takes advantage of the \emph{differential savings}
and \emph{recursive savings}, and can greatly reduce the number of
additions in calculating $\mathbf{M}\mathbf{x}$, although the reduced
additive complexity, denoted by
$\mathcal{C}_\mathrm{CSE}(\mathbf{M})$, is not guaranteed to be the
minimum. Like other CSE algorithms, the CSE algorithm in
\cite{Chen2009c} is randomized, and the reduction results of different
runs are not necessarily the same. Therefore in practice, a better
result can be obtained by first running the CSE algorithm many times
and then selecting the smallest number of additions. The CSE algorithm
in \cite{Chen2009c} greatly reduces the additive and overall
complexities of CFFTs with lengths up to $1023$, but it is much more
difficult to reduce the additive complexities of longer CFFTs. This is
because though the CSE algorithm in \cite{Chen2009c} has a polynomial
complexity (it is shown that its complexity is $O(N^4+N^3M^3)$), the
runtime and memory requirements become prohibitive when $M$ and $N$
are very large, which occurs for long CFFTs.

\subsection{Prime-Factor and Cooley-Turkey Algorithms}
Both the prime-factor algorithm and Cooley-Turkey algorithm first
decompose an $N$-point DFT into shorter sub-DFTs, and then construct
the $N$-point DFT from the sub-DFTs \cite{BlahutFast}. The
prime-factor algorithm requires that the length $N$ has at least two
co-prime factors, i.e., there exist two co-prime numbers $N_1$ and
$N_2$ such that $N=N_1N_2$.  For an integer $i \in \{0, 1, \cdots,
N-1\}$, there is a unique integer pair $(i_1, i_2)$ such that $0\le
i_1 \le N_1-1$, $0\le i_2 \le N_2-1$, and $i=i_1N_2+i_2N_1\pmod{N},$
since $N_1$ and $N_2$ are co-prime. For any integer $j\in\{0,1,\cdots,
N-1\}$, let $j_1 = j \pmod{N_1},\;\;\;j_2=j\pmod{N_2},$ where $0\le
j_1 \le N_1-1$ and $0 \le j_2 \le N_2-1$.  By Chinese remainder
theorem, $(j_1, j_2)$ uniquely determines $j$, and $j$ can be
represented by $j=j_1N_2^{-1}N_2+j_2N_1^{-1}N_1\pmod{N}$, where
$N_2^{-1}N_2=1\pmod{N_1}$ and $\;N_1^{-1}N_1=1\pmod{N_2}.$
Substituting the above representation of $i$ and $j$
in \eqref{eq:DFT}, we get
$\alpha^{ij}=(\alpha^{N_2})^{i_1j_1}(\alpha^{N_1})^{i_2j_2},$
where $\alpha^{N_2}$ and $\alpha^{N_1}$ are the $N_1$-th root and
$N_2$-th root of 1, respectively. Therefore, \eqref{eq:DFT} becomes
\begin{equation}
  \label{eq:PFA}
  F_j = 
  \underbrace{\sum_{i_1=0}^{N_1-1}\Big(\overbrace{\sum_{i_2=0}^{N_2-1}f_{i_1N_2+i_2N_1}\alpha^{N_1i_2j_2}}^{N_2-\mathrm{point\,DFT}}\Big)\alpha^{N_2i_1j_1}}_{N_1-\mathrm{point\,DFT}}.
\end{equation}
In this way, the $N$-point DFT is obtained by using $N_1$- and
$N_2$-point sub-DFTs. The $N$-point DFT result is derived by first
carrying out $N_1$ $N_2$-point DFTs and $N_2$ $N_1$-point DFTs, and
then combining the results according to the representation of $j$. The
prime-factor algorithm can also be applied to $N_1$- and $N_2$-point
DFTs if they have co-prime factors.

The Cooley-Turkey algorithm has a different decomposition strategy
from the prime-factor algorithm.  Let $N=N_1N_2$, where $N_1$ and
$N_2$ do not have to be co-prime. Let $i=i_1+i_2N_1$, where $0\le i_1
\le N_1-1$ and $0\le i_2 \le N_2-1$, and $j=j_1N_2+j_2$, where $0\le
j_1\le N_1-1$ and $0\le j_2 \le N_2-1$.  Then \eqref{eq:DFT} becomes
\begin{equation}
  \label{eq:CTA}
  F_j
  =\underbrace{\sum_{i_1=0}^{N_1-1}\Big(\overbrace{\sum_{i_2=0}^{N_2-1}f_{i_1+i_2N_1}\alpha^{N_1i_2j_2}}^{N_2-\mathrm{point\,DFT}}\Big)\alpha^{i_1j_2}\alpha^{N_2i_1j_1}}_{N_1-\mathrm{point\,DFT}}.
\end{equation}
In this way, the Cooley-Turkey algorithm also decomposes the $N$-point
DFT into $N_1$- and $N_2$-point DFTs. However, compared with
\eqref{eq:PFA}, \eqref{eq:CTA} has an extra term $\alpha^{i_1j_2}$,
which is called \emph{twiddle factor} and incurs additional
multiplicative complexity. The Cooley-Turkey algorithm can be used for
arbitrary non-prime length $N$, including the prime powers to which
case the prime-factor algorithm cannot be applied. The Cooley-Turkey
algorithm is very suitable if $N$ has a lot of small factors, for
example, $2^n$-point DFT by the Cooley-Turkey algorithm requires
$O(n\cdot 2^n)$ multiplications.


\section{$p$-point Cyclic Convolutions over GF$(2^{m})$}
\label{sec:conv}
Efficient short cyclic convolution algorithms play an essential role
in the multiplicative complexity reduction of CFFTs. Note the lengths
of cyclic convolutions involved in CFFTs are the same as the sizes of
the conjugate classes. Since the sizes of all possible conjugate
classes in GF$(2^{m})$ are divisors of $m$, efficient algorithms for
only short cyclic convolutions are needed, since they determine the
multiplicative complexities of CFFTs.

Despite their significance, there is no general algorithms for
efficient cyclic convolutions of arbitrary length over finite fields.
Of course, efficient ad hoc algorithms for 2- to 9-point cyclic
convolution can be found in the literature (4- and 8-point can be
found in \cite{private,Afanasyev1987,Churkov2004}, and their details
are included in Appendix \ref{sec:convs} due to their limited access,
and the rest can be found in \cite{BlahutFast} and
\cite{BlahutECC}). Furthermore, cyclic convolutions with composite
length can be constructed with multi-dimensional technology described
in \cite{BlahutFast}. For instance, 10-point cyclic convolution
algorithms can be constructed based on 2- and 5-point algorithms,
while 12-point cyclic convolution algorithm is constructed based on 3-
and 4-point algorithms. However, an efficient algorithm for cyclic
convolutions of larger prime length (for example, 11- or 13-point) is
not available in the open literature. We can implement these cyclic
convolutions via the convolution theorem. Although the DFTs and IDFT
can be implemented by the Winograd algorithm \cite{Winograd1978} or
the Rader algorithm \cite{Rader1968}, this approach remains
inefficient, especially for small to moderate lengths. In
\cite{WM_IT83}, strategies to derive cyclic convolution algorithms
directly over any finite field GF$(q^m)$ were developed.
Unfortunately, these methods are applicable only to lengths $q^m-1$ or
their factors.

Herein for an odd prime $p$, we reformulate a $p$-point cyclic
convolution as a product of a $(p-1)\times (p-1)$ Toeplitz matrix and
a vector. Since $p-1$ is composite, this product can be readily
obtained by multi-dimensional technology from well-known TMVP of very
small sizes, leading to efficient bilinear algorithms for $p$-point
cyclic convolutions. Since these cyclic convolutions will be used for
CFFTs over GF$(2^l)$, we focus on cyclic convolutions over
GF$(2^l)$. However, our reformulation can be readily extended to the
real and complex fields as well as more general finite
fields. Furthermore, by multi-dimensional technology, we can also
obtain efficient algorithms for $p^n$-point cyclic convolutions. These
algorithms are also key to long CFFTs.

For a $p$-dimensional vector $\mathbf{x}=(x_0, x_1, \cdots,
x_{p-1})^T$ over some field GF$(2^l)$, where $p$ is any odd prime
integer, we consider its corresponding polynomial
$X(w)=\sum_{i=0}^{p-1}x_iw^i$. Assuming that the $p$-point cyclic
convolution of two vectors $\mathbf{x}$ and $\mathbf{y}$ is
$\mathbf{z}$, all of which are $p$-dimensional vectors over GF$(2^l)$,
their corresponding polynomials are related by \cite{BlahutFast}
\begin{equation}
  \label{eq:polyconv}
  Z(w)=X(w)Y(w)\pmod{w^p+1}.
\end{equation}
Note that $w^p+1=(w+1)(w^{p-1}+w^{p-2}+\cdots + 1)$, and $w+1$ and
$w^{p-1}+w^{p-2}+\cdots + 1$ are co-prime in GF$(2^l)$. Hence by
Chinese remainder theorem, $Z(w)$ can be uniquely determined by $Z_0$
and $Z'(w)=\sum_{i=0}^{p-2}Z'_iw^i$, where
\begin{equation}
  \label{eq:CRT}
  \begin{split}
    Z_0&=Z(w)\pmod{w+1},\\
    Z'(w)&=Z(w)\pmod{w^{p-1}+w^{p-2}+\cdots +1}.
  \end{split}
\end{equation}
It is easy to see that $Z_0 = \sum_{i=0}^{p-1}z_i$,
$Z'_i=z_i+z_{p-1}$, and the vector $\mathbf{Z}^\dagger = (Z_0, Z'_0,
Z'_1, \cdots, Z'_{p-2})^T$ can be derived by multiplying the vector
$\mathbf{z}$ with an $p\times p$ matrix $\mathbf{B}$ with structure
$$
\mathbf{B} =
\begin{bmatrix} 1 & 1 &  \dotso & 1 \\
  & & & 1\\
  \multicolumn{3}{c}{\mathbf{I}_{p-1}} & \vdots\\
  & & & 1
\end{bmatrix}
$$
where $\mathbf{I}_{p-1}$ is a $(p-1)\times (p-1)$ identity
matrix. That is, $\mathbf{Z}^\dagger=\mathbf{B}\mathbf{z}$.

To compute the $p$-point cyclic convolution of $\mathbf{x}$ and
$\mathbf{y}$, we first compute
$\mathbf{X}^\dagger=\mathbf{B}\mathbf{x}$ and
$\mathbf{Y}^\dagger=\mathbf{B}\mathbf{y}$, then compute
$\mathbf{Z}^\dagger$ from $\mathbf{X}^\dagger$ and
$\mathbf{Y}^\dagger$, and finally,
$\mathbf{z}=\mathbf{B}^{-1}\mathbf{Z}^\dagger$.  With the same
partitioning scheme aforementioned and equations \eqref{eq:polyconv}
and \eqref{eq:CRT}, it is easy to see that $Z_0=X_0Y_0$, and
\begin{equation}
  \label{eq:Zprime}
  Z'(w)=X'(w)Y'(w) \pmod{w^{p-1}+w^{p-2}+\cdots+1},
\end{equation}
and hence we can compute
$\mathbf{Z}^\dagger=(Z_0, \mathbf{Z}'^T)^T$.

From \eqref{eq:Zprime}, the polynomial product can be computed as
\begin{align}
  \label{eq:polyprod}
  X'(w)Y'(w)&=\sum_{k=0}^{p-2}\sum_{j=0}^{p-2}(Y'_{k-j}+Y'_{k-j+p}+Y'_{p-1-j})X'_jw^k\nonumber\\
  &\pmod{w^{p-1}+w^{p-2}+\cdots+1},
\end{align}
and hence the vector $\mathbf{Z}'$ can be computed through a matrix
product $\mathbf{Z}'=\mathbf{M}\mathbf{X}'$, where the elements of
matrix $\mathbf{M}$ are
\begin{equation}
  \label{eq:Melem}
  M_{k,j}=Y'_{k-j}+Y'_{k-j+p}+Y'_{p-1-j}.
\end{equation}
Note that in \eqref{eq:polyprod} and \eqref{eq:Melem}, $Y'_i$ are considered as zero outside its
valid range, i.e., $Y_i'=0$ if $i<0$ or $i>p-2$.

We can check that $\mathbf{B}$ is an invertible matrix, and
$\mathbf{B}^{-1}$ is given by
$$
\mathbf{B}^{-1}=
\begin{bmatrix}
  1 &\mathbf{A}_1\\
  \mathbf{A}_2 & \mathbf{A}_3
\end{bmatrix}
$$
where the length-$(p-1)$ row vector $\mathbf{A}_1=(0, 1, 1,
\cdots, 1)$, the length-$(p-1)$ column vector $\mathbf{A}_2=(1, 1,
\cdots, 1)^T$, and $(p-1)\times (p-1)$ matrix $\mathbf{A}_3$ has $0$
on the first upper diagonal and $1$ everywhere else.

Now consider the product of $\mathbf{B}^{-1}$ and a length-$p$ column
vector $\mathbf{U}$:
$$\mathbf{B}^{-1}\begin{bmatrix}U_0\\\mathbf{U}'\end{bmatrix} =
  \begin{bmatrix} 1 & \mathbf{A}_1\\
  \mathbf{A}_2 & \mathbf{A}_3
\end{bmatrix}
\begin{bmatrix}
  U_0\\ \mathbf{U}'
\end{bmatrix} = \begin{bmatrix}V_0\\\mathbf{V}'
\end{bmatrix}$$ where $U_0$, $\mathbf{U}'$, and $V_0$, $\mathbf{V}'$
are appropriate partitions of the vector $\mathbf{U}$ and the
multiplication result vector $\mathbf{V}$, respectively.  Values of
$V_0$ and $\mathbf{V}'$ can be computed as $V_0 = U_0 + \mathbf{A}_1
\mathbf{U}'$ and $\mathbf{V}' = \mathbf{A}_2 U_0 + \mathbf{A}_3
\mathbf{U}'$.  Note that $\mathbf{A}_1$ and $\mathbf{A}_3$ are related
as $\mathbf{A}_1 = (1, 1, \dotsc, 1)\,\mathbf{A}_3$.  This implies
that the sum of the components of $\mathbf{A}_3 \mathbf{U}'$ gives
$\mathbf{A}_1 \mathbf{U}'$.  Furthermore, $\mathbf{A}_2$ contains only
1's.  Thus the computation of $V_0$ and $\mathbf{V}'$ reduces to
\begin{equation}
	\label{V1_V2}
\begin{split}
  V_0 &= U_0 + \sum(\mathbf{A}_3 \mathbf{U}') \\
  \mathbf{V}' &= [U_0, U_0, \dotsc, U_0]^\mathrm{T} +
  \mathbf{A}_3 \mathbf{U}'.
\end{split}
\end{equation}
Eq. \eqref{V1_V2} shows that multiplying a vector with
$\mathbf{B}^{-1}$ needs only an evaluation of $\mathbf{A}_3
\mathbf{U}'$.

The cyclic convolution result $\mathbf{z}$ is obtained by first
multiplying $\mathbf{A}_3$ and $\mathbf{Z}'$. Thus one need to compute
$\mathbf{R}\mathbf{X}'$ where the $(p-1)\times(p-1)$ matrix
$\mathbf{R}=\mathbf{A}_3\mathbf{M}$. We now show by direct computation
that $\mathbf{R}$ is a Toeplitz matrix.  From the structure of
$\mathbf{A}_3$, we have
\begin{equation}
\label{eq:Relem}
R_{i,j} = M_{i+1,j}+\sum_{k=0}^{p-2} M_{k, j}.
\end{equation}
>From \eqref{eq:Melem}, using appropriate ranges for the three terms we
get
\begin{equation}
\label{eq:M-ki-sum}
\begin{split}
\sum_{k=0}^{p-2} M_{k, j} = Y'_{p-1-j}+ \sum_{s=0}^{p-2} Y'_s.
\end{split}
\end{equation}
Finally, combining \eqref{eq:Melem}, \eqref{eq:Relem} and
\eqref{eq:M-ki-sum} gives
\begin{equation}
	R_{i, j} = Y'_{i-j+1} + Y'_{i-j+p+1} +\sum_{s=0}^{p-2} Y'_s.
\label{eq:Relem2}
\end{equation}
Since $R_{i,j}$ is a function of only $i-j$, $\mathbf{R}$ is a
Toeplitz matrix.
Recall that $Y'_i$ is assumed zero if its index is outside the valid
range from 0 to $p-2$.  Thus in \eqref{eq:Relem2}, at most one of the
first two terms is valid for any combination of $i$ and $j$.

\begin{figure}[hbtp]
  \centering
  \includegraphics[width=\columnwidth]{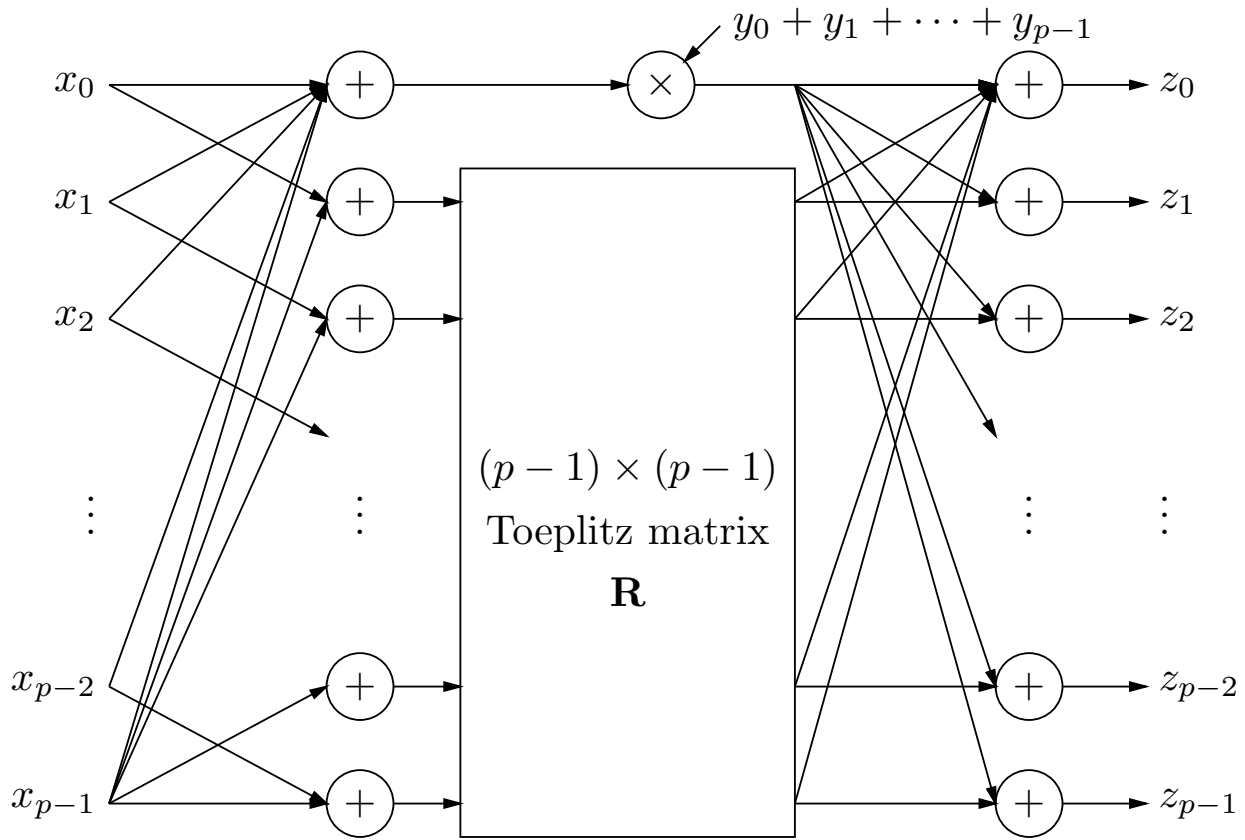}
  \caption{$p$-point cyclic convolution.}
  \label{fig:convp}
\end{figure}

Fig.~\ref{fig:convp} illustrates our algorithm for $p$-point cyclic
convolutions, which relies on the implementation of
$\mathbf{R}\mathbf{X}'$. Direct implementation of
$\mathbf{R}\mathbf{X}'$ requires $(p-1)^2$ multiplications, but we can
reduce it since $\mathbf{R}$ is a Toeplitz matrix.  For any odd prime
$p>3$, $p-1$ is composite and $\mathbf{R}\mathbf{X}'$ can be obtained
by using multi-dimensional technology from TMVP of smaller sizes
\cite{Allwright1971,Pitassi1971,Agarwal1974,Winograd1980,Fan2007}. For
example, CFFTs over GF$(2^{11})$, GF$(2^{13})$, GF$(2^{17})$, and
GF$(2^{19})$, involve $11$-, $13$-, $17$-, and $19$-point cyclic
convolutions, respectively. Using our reformulations, these cyclic
convolutions can be obtained from a TMVP of $2\times 2$, $3\times 3$,
and $5\times 5$, which are provided in Appendix~\ref{sec:TMVP}. Hence
our reformulation leads to efficient cyclic convolution algorithms for
odd prime $p$ for $p\leq 19$, which are sufficient for all CFFTs over
characteristic-$2$ fields as large as GF$(2^{19})$.

This reformulation is also applicable to a prime greater than $19$,
where $p-1$ may have a prime factor $p'$ greater than five. In this
case, one can use two ad hoc techniques to proceed. First, one can
break a $p'\times p'$ matrix into blocks, and treat them
separately. Second, one can extend the $p'\times p'$ matrix to a
larger matrix so that it remains a Toeplitz matrix and its size
becomes composite again. The complexities of cyclic convolution
algorithms obtained through this reformulation are much smaller than
direct implementation. For example, we can first extend the
$(p-1)\times (p-1)$ Toeplitz matrix to a $2^{\lceil \log_2(p-1)\rceil}
\times 2^{\lceil \log_2(p-1)\rceil}$ matrix, and it requires fewer
than $3^{\lceil \log_2(p-1)\rceil}$ multiplications if we use the
two-way split method described in \cite{Fan2007}.

We note that a $p$-point cyclic convolution can be formulated as a
$p\times p$ circulant matrix vector product. Since a circulant matrix
is a special case of Toeplitz matrix, one can of course apply the two
ad hoc techniques described above to this $p\times p$ Toeplitz matrix
directly. However, since our reformulation turns a $p$-point cyclic
convolution into a $(p-1) \times (p-1)$ TMVP, which directly benefit
from multi-dimensional technologies, at the expense of only one extra
multiplication, we believe our reformulation will lead to lower
multiplicative complexity. We cannot prove this analytically, but will
illustrate this point below with an example.

We also remark that our reformulation leads to bilinear algorithms for
cyclic convolutions, which can be implemented efficiently since the
pre- and post-addition matrices are all binary.




\subsection{Example: 11-point Convolution Algorithm over GF$(2^m)$}
\label{sec:11conv}
To illustrate the advantages of our reformulation above, we derive our
efficient $11$-point cyclic convolution algorithm over GF$(2^{11})$
and compare its multiplicative complexity with some other approaches.
By using well-known $2\times 2$ and $5\times 5$ TMVP, we obtain an
$11$-point cyclic convolution algorithm
$\mathbf{z}=\mathbf{Q}^{(11)}(\mathbf{R}^{(11)}\mathbf{y}\cdot
\mathbf{P}^{(11)}\mathbf{x})$, where the matrices $\mathbf{Q}^{(11)}$,
$\mathbf{P}^{(11)}$, and $\mathbf{R}^{(11)}$ are given in
Appendix~\ref{sec:convs}. Since the $10\times 10$ TMVP requires $42$
multiplications, our 11-point cyclic convolution requires $43$
multiplications.

Let us compare this multiplicative complexity with the two ad hoc
techniques.  First, we can partition the $11\times 11$ circulant
matrix into a $10\times 10$ Toeplitz matrix, a $10\times 1$ column
vector, a $1 \times 10$ row vector, and a single element, and then
apply the multi-dimensional technology to the $10\times 10$ TMVP. In
addition to the $10\times 10$ TMVP, this approach requires $21$
extra  multiplications, as opposed to one in our approach.
Second, we can extend the $11\times 11$ circulant matrix to a
$12\times 12$ Toeplitz matrix, and then apply the multi-dimensional
technology to this matrix. A $12\times 12$ TMVP requires $54=3\times 3
\times 6$ multiplications. Taking into account that we pad a zero to
the $11\times 1$ vector and that the last element of the TMVP is not
needed, two multiplications can be saved, and we need $52$
multiplications in total (note that this total multiplicative
complexity is the same regardless of the order of decomposition of
$12$).  We can also extend the $11\times 11$ circulant matrix to a
$15\times 15$ Toeplitz matrix or a $16\times 16$ one, which require
$66$ and $60$ multiplications, respectively.  Our reformulation is
more efficient than these ad hoc techniques in terms of the
multiplicative complexity. This is because our reformulation turns a
$p$-point cyclic convolution into a $(p-1) \times (p-1)$ TMVP, which
directly benefit from multi-dimensional technologies, at the expense
of only one extra multiplication.

We also compare our result with the implementation via convolution
theorem, i.e., first multiply the DFTs of the two vector
component-wisely, and then compute the inverse DFT of the resulting
vector. If we use the Rader's algorithm to implement the DFT and
inverse DFT, it needs 101 multiplications in total. Hence this
approach is less efficient than ours.

By using the CSE algorithm in \cite{Chen2009c}, our $11$-point cyclic
convolution algorithm requires $43$ multiplications and $164$
additions. When we use this algorithm in CFFTs over GF$(2^{11})$, one
of the two inputs is known in advance. Our algorithm requires $42$
multiplications since one of the multiplication has an operand of one,
and $120$ additions because the additions involving the known input
can be pre-computed.

\section{Long Cyclotomic Fourier Transforms}
\label{sec:longCFFT}

\subsection{2047-point CFFT over GF$(2^{11})$}
\label{sec:2047}

The efficient algorithm for 11-point cyclic convolution we designed in
\ref{sec:11conv} is the key to
the CFFTs over GF$(2^{11})$.
Direct implementation of $2047$-point CFFT with this cyclic
convolution algorithm requires $7812$ multiplications and $2130248$
additions. The prohibitively high additive complexity is dominated by
the multiplication between the $2047 \times 2047$ matrix $\mathbf{A}$
and a 2047-dimensional vector, which requires $2095280$ additions.
Unfortunately, if we use the CSE algorithm in \cite{Chen2009c} to
reduce its additive complexity, the time complexity of the CSE
algorithm itself is too high (it needs months to finish).

Due to the high time complexity of the CSE algorithm in
\cite{Chen2009c}, we have tried a simplified CSE algorithm with
limited success. In the original CSE algorithm in \cite{Chen2009c},
only one of the patterns with the greatest recursive savings is
selected and removed in one round of iteration.  Instead of selecting
only one pattern, our simplified CSE algorithm has a reduced time
complexity as it removes multiple patterns at one time. The reduced
time complexity of the simplified CSE algorithm allows us to reduce
the additive complexity for the $2047$-point CFFT to $529720$
additions, about one fourth of that for the direct
implementation. Despite this improvement, the effectiveness of this
simplified CSE algorithm is rather limited.

\subsection{Difficulty with Long CFFTs}
Consider an $N$-point CFFT over GF$(2^l)$. Suppose $C_{s_0}, C_{s_1},
\cdots, C_{s_{m-1}}$ are $m$ cyclotomic cosets modulo $N$ over
GF$(2)$, and $|C_{s_k}|=m_k$. Suppose an $m_k$-point cyclic
convolution can be done with $\mathcal{M}(m_k)$ multiplications, and
hence implementing the $N$-point DFT with the CFFT directly requires
$\sum_{k=0}^{m-1}\mathcal{M}(m_k)$ multiplications and
$\mathcal{C}(\mathbf{AQ})+\mathcal{C}(\mathbf{P})$ additions, where
$\mathcal{C}(\cdot)$ denotes the number of additions we need to
evaluate the product of a binary matrix and a vector. The
multiplicative complexity can be further reduced because we can
pre-compute the vector $\mathbf{c}$ in \eqref{eq:CFFT} and some of its
elements may be unitary. Then the CSE algorithm can be applied to the
matrices $\mathbf{AQ}$ and $\mathbf{P}$ to reduce
$\mathcal{C}(\mathbf{AQ})$ and $\mathcal{C}(\mathbf{P})$ to
$\mathcal{C}_\mathrm{CSE}(\mathbf{AQ})$ and
$\mathcal{C}_\mathrm{CSE}(\mathbf{P})$, respectively. Since
$\mathbf{P}=\mathrm{diag}(\mathbf{P}_0, \mathbf{P}_1, \cdots,
\mathbf{P}_{m-1})$ is a block diagonal matrix, we have
$\mathcal{C}_\mathrm{CSE}(\mathbf{P})=\sum_{i=0}^{m-1}\mathcal{C}_\mathrm{CSE}(\mathbf{P}_i)$. Therefore,
we can reduce the additive complexity of each $\mathbf{P}_i$ to get a
better result of $\mathcal{C}(\mathbf{P})$. Since the size of
$\mathbf{P}_i$ is much smaller than that of $\mathbf{P}$, it allows us
to run the CSE algorithm many times to achieve a smaller additive
complexity. However, the matrix $\mathbf{AQ}$ is not a block diagonal
matrix, and therefore we have to apply the CSE algorithm directly to
$\mathbf{AQ}$. When the size of $\mathbf{AQ}$ is large, the CSE
algorithm in \cite{Chen2009c} requires a lot of time and memory and
hence it is impractical for extremely long DFTs. As mentioned above,
it would take months for the CSE algorithm in \cite{Chen2009c} to
reduce the additive complexity of $2047$-point CFFT over GF$(2^{11})$,
let alone $4095$-point CFFTs over GF$(2^{12})$. The prohibitively high
time complexity of the CSE algorithm in \cite{Chen2009c} and the
limited effectiveness of the simplified CSE algorithm motivate our
composite cyclotomic Fourier transforms.

\section{Composite Cyclotomic Fourier Transforms}
\label{sec:CCFT}

\subsection{Composite Cyclotomic Fourier Transforms}
\label{sec:factors}
Instead of simplifying the CSE algorithm or designing other low
complexity optimization algorithms, we propose composite cyclotomic
Fourier transforms by first decomposing a long DFT into shorter
sub-DFTs, via the prime-factor or Cooley-Turkey algorithms, and then
implementing the sub-DFTs by CFFTs. Note that both the decompositions
require only that $\alpha$ is a primitive $N$-th root of 1, hence they
can be extended to finite fields easily.  When $N$ is prime, our CCFTs
reduce to CFFTs. When $N$ is composite, we first decompose the DFT
into shorter sub-DFTs, and then combine the sub-DFT results according
to \eqref{eq:PFA} or \eqref{eq:CTA}. The shorter sub-DFTs are
implemented by CFFTs to reduce their multiplicative complexities, and
then we use the CSE algorithm in \cite{Chen2009c} to reduce their
additive complexities. Finally, when $N$ has multiple factors, the
factorization can be carried out recursively.

Suppose the length of the DFT is composite, i.e., $N=N_1N_2$. Either
the prime-factor or the Cooley-Turkey algorithms can be used to
decompose the $N$-point DFT into sub-DFTs when $N_1$ and $N_2$ are
co-prime.  When $N_1$ and $N_2$ are not co-prime, only the
Cooley-Turkey algorithm can be used. It is easy to show that if $N_1$
and $N_2$ are co-prime, the prime-factor and Cooley-Turkey algorithms
lead to the same additive complexity for CCFTs, but the Cooley-Turkey
algorithm results in a higher multiplicative complexity due to the
extra multiplications of twiddle factors. Hence the prime-factor
algorithm is better than the Cooley-Turkey algorithm in this case, and
the Cooley-Turkey algorithm is used only if the prime-factor algorithm
cannot be applied.

We denote the multiplicative and additive complexity of an
$N$-point DFT by $\mathcal{K}^\mathrm{mult}(N)$ and
$\mathcal{K}^\mathrm{add}(N)$, respectively, and the algorithm used to
implement this DFT is specified in the subscription of
$\mathcal{K}$. Suppose $N=\prod_{i=1}^sN_i$, and the total number of
non-unitary twiddle factors required by the Cooley-Turkey algorithm
decompositions is denoted by $T$,
then the complexity of this decomposition is given by
\begin{align}
  \label{eq:additive}
  \mathcal{K}^\mathrm{add}_\mathrm{CCFT}(N)&=\sum_{i=1}^s\frac{N}{N_i}\mathcal{K}^\mathrm{add}_\mathrm{CFFT}(N_i),\\
  \label{eq:multiplicative}
  \mathcal{K}^\mathrm{mult}_\mathrm{CCFT}(N)&=\sum_{i=1}^s\frac{N}{N_i}\mathcal{K}^\mathrm{mult}_\mathrm{CFFT}(N_i)+T.
\end{align}
For $N|2^{l}-1$ for $4 \le l \le 12$, there
is at most one pair of $N_i$'s that are not co-prime in the
decomposition of $N$, say $N_1$ and $N_2$,  without loss of
generality. In this case, $T=\frac{N}{N_1N_2}(N_1-1)(N_2-1)$. If all
the elements in the decomposition of $N$ are co-prime to each other,
then $T=0$.

The decomposition allows our CCFTs to achieve low complexities for
several reasons. First, this divide-and-conquer strategy is used in
many fast Fourier transforms. If we assume CFFTs have quadratic
additive complexities with their length $N$ when directly implemented
(this assumption is at least supported by the additive complexities of
the CFFTs without CSE in Table \ref{tab:comp}),
the CCFT decomposition reduces the additive complexity from $O(N^2)$
to $O(N\sum_{i=1}^sN_i)$.  Second, the lengths of the sub-DFTs are
much shorter, which enables us to apply several powerful but
complicated techniques to reduce the complexities of the sub-DFTs. For
example, it takes much less time and memory to apply the CSE algorithm
in \cite{Chen2009c} to the sub-DFTs, and thus we can run it multiple
times to get a better reduction result. Third, when the length of the
DFT admits different factorizations (for example, $2^6-1=63=3\times 21
= 9 \times 7$), we choose the decomposition(s) with the lowest
complexity.



\subsection{Complexity Reduction}
\label{sec:result}
We reduce the additive complexities of our CCFTs in three
steps. First, we reduce the complexities of short cyclic
convolutions. Second, we use these short cyclic convolutions to
construct CFFTs of moderate lengths. Third, we use CFFTs of moderate
lengths as sub-DFTs to construct our CCFTs.

Efficient short cyclic convolution algorithms are the keys to the
multiplicative complexity reduction of CFFTs and our CCFTs, and hence
our \textbf{first} step is to reduce the computational complexities of
small size cyclic convolutions. Suppose an $L$-point cyclic
convolution $\mathbf{b}^{(L)}\otimes \mathbf{a}^{(L)}$ is calculated
with the bilinear form
$\Q^{(L)}(\mathbf{R}^{(L)}\mathbf{b}^{(L)}\cdot\mathbf{P}^{(L)}\mathbf{a}^{(L)})$.
Since $\mathbf{b}^{(L)}$ is the normal basis in our CCFTs,
$\mathbf{R}^{(L)}\mathbf{b}^{(L)}$ can be precomputed to reduce
multiplicative complexity.
We apply the CSE algorithm in \cite{Chen2009c} to reduce the additive
complexities in the multiplication with binary matrices
$\mathbf{Q}^{(L)}$ and $\mathbf{P}^{(L)}$. The complexity reduction
results $\mathcal{C}_\mathrm{CSE}(\Q^{(L)})$,
$\mathcal{C}_\mathrm{CSE}(\mathbf{P}^{(L)})$, the total additive
complexity
$\mathcal{C}_\mathrm{CSE}(\mathbf{Q}^{(L)})+\mathcal{C}_\mathrm{CSE}(\mathbf{P}^{(L)})$,
and the multiplicative complexities are listed in Table
\ref{tab:convCmplx}.

\begin{table}[!htb]
  \centering
  \caption{Complexities of short cyclic convolutions over GF$(2^l)$.}
  \begin{tabular}{c|c|ccc}
    \hline
     \multirow{2}{*}{$L$} &  \multirow{2}{*}{mult. }  & \multicolumn{3}{c}{additive complexities} \\ \cline{3-5}
     & & $\mathcal{C}_\mathrm{CSE}(\Q^{(L)})$ &
    $\mathcal{C}_\mathrm{CSE}(\mathbf{P}^{(L)})$ & total\\
    \hline
    2  & 1  & 2   & 1  & 3\\
    3  & 3  & 5   & 4  & 9\\
    4  & 5  & 9   & 4  & 13\\
    5  & 9  & 16  & 10 & 26\\
    6  & 10 & 21  & 11 & 32\\
    7  & 12 & 24  & 23 & 47\\
    8  & 19 & 35  & 16 & 51\\
    9  & 18 & 40  & 31 & 71\\
    10 & 28 & 52  & 31 & 83\\
    11 & 42 & 76  & 44 & 120\\
    12 & 32 & 53  & 34 & 87\\
    \hline
  \end{tabular}
  \label{tab:convCmplx}
\end{table}

The \textbf{second} step is to reduce the additive complexity of CFFTs with moderate
lengths, which will be used to build our CCFTs. Their moderate lengths allow us to use multiple techniques to reduce their additive complexities.
\begin{itemize}
\item First, for any CFFT, we run the CSE algorithm in
  \cite{Chen2009c} multiple times and then choose the best results.
\item Second, for each CFFT in (\ref{eq:CFFT}), we may reduce
  $\mathcal{C}(\mathbf{AQ})$ together as a whole, or reduce
  $\mathcal{C}(\mathbf{A})$ and $\mathcal{C}(\mathbf{Q})$
  separately. Since $(\mathbf{AQ})\mathbf{v}=\mathbf{A}(\mathbf{Qv})$,
  $\mathcal{C}_\mathrm{opt}(\mathbf{AQ})\le\mathcal{C}_\mathrm{opt}(\mathbf{A})+\mathcal{C}_\mathrm{opt}(\mathbf{Q})$. However,
  this property may not hold for the CSE algorithm because the CSE
  algorithm may not
  find the optimal solutions. Furthermore, we may benefit from
  reducing $\mathcal{C}(\A)$ and $\mathcal{C}(\Q)$ separately for the
  following reasons. First, $\mathbf{\Q}$ has a block diagonal
  structure, which is similar as $\mathbf{P}$, therefore we can find a
  better reduction result for $\mathcal{C}(\Q)$. Second, $\mathbf{AQ}$
  has much more columns than $\mathbf{A}$, and hence the CSE algorithm
  requires less memory and time to reduce $\mathbf{A}$ than to reduce
  $\mathbf{AQ}$.
\item Third, there is flexibility in terms of normal bases used to
  construct the matrix $\mathbf{A}$ in (\ref{eq:CFFT}), and this
  flexibility can be used to further reduce the additive complexity of
  any CFFT.  For each cyclotomic coset, a normal basis is needed.  A
  normal basis is not unique in finite fields, and any normal basis
  can be used in the construction of the matrix $\mathbf{A}$, leading
  to the same multiplicative complexity. But different normal bases
  result in different $\mathbf{A}$ and hence different additive
  complexities due to $\mathbf{A}$.
  There are several options regarding the normal basis. One can simply
  choose a fixed normal basis for all cyclotomic cosets of the same
  size as in \cite{Chen2009c}.  A more ideal option is to enumerate
  all possible normal bases and their corresponding $\mathbf{A}$ and
  to select the smallest additive complexity. However, when the
  underlying field is large,
  the number of possible normal basis is very large, and hence it
  becomes infeasible to enumerate all possible constructions.  Thus,
  in this paper we use a compromise of these two options: for each
  cyclotomic coset we choose a normal basis at random and the
  combination of random normal bases leads to $\mathbf{A}$; we
  minimize the complexity over as many combinations as complexity
  permits. We refer to this as a random normal basis option.
\end{itemize}
We emphasize that all three techniques require multiple runs of the
CSE algorithm. Since the time and memory requirements of the CSE
algorithm grows with the length of DFT, the moderate lengths of the
sub-DFTs is the key enabler of these techniques.

For any $k \le 320$ so that $k | 2^l-1$ ($4\le l \le 12$), the
multiplicative and additive complexities of the $k$-point CFFT are
shown in Table \ref{tab:shortFFTs}. Table \ref{tab:shortFFTs} shows four
different schemes to reduce the additive complexity for CFFTs. Schemes
A and B both use the fixed normal basis option in the construction of
the matrix $\mathbf{A}$, while schemes C and D are based on the random
normal basis option. Schemes A and C reduce $\mathcal{C}(\mathbf{A})$
and $\mathcal{C}(\mathbf{Q})$ separately, while schemes B and D
reduces $\mathcal{C}(\mathbf{AQ})$ as a whole. For smaller CFFTs, we
typically minimize the complexity over hundreds of combinations of
normal bases, and fewer combinations for longer CFFTs.
In Table \ref{tab:shortFFTs}, the smallest additive complexities are
in boldface font.
We observe that the random normal basis option offers further additive
complexity reduction in most of the cases. However, since the fixed
normal basis is not necessarily one of the combinations, in some cases
the fixed normal basis option outperforms the random normal basis
option. Also, sometimes applying the CSE to $\mathbf{AQ}$
together as a whole leads to lower complexity, and in some cases it is
better to apply the CSE to $\mathbf{A}$ and
$\mathbf{Q}$ separately.

\begin{table}[!htb]
  \centering
  \caption{The complexities of the CFFTs whose lengths are less
    than  $320$ and are factors of $2^l-1$ for $1\le l \le 12$.}
  \begin{tabular}{l|l|c|cccc}
    \hline
    \multirow{2}{*}{$N$} & \multirow{2}{*}{$l$} &
    \multirow{2}{*}{mult.} & \multicolumn{4}{c}{additive complexities}\\
    \cline{4-7}
    & & & A  & B & C & D\\
    \hline
    3 & 2 & 1 & \textbf{6} & \textbf{6} & \textbf{6} & \textbf{6}  \\
    5 & 4 &5 & 20 & \textbf{16} & 20 & \textbf{16} \\
    7 & 3 &6 & 31 & \textbf{24} & 31 & \textbf{24} \\
    9 & 6 &11 & 51 & \textbf{48} & 51 & \textbf{48} \\
    11 & 10 &28 & 109 & 102 & 102 & \textbf{84} \\
    13 & 12 &32 & 125 & 100 & 110 & \textbf{91} \\
    15 & 4  &16 & 87 & \textbf{74} & 87 & \textbf{74} \\
    17 & 8  &38 & 153 & 163 & \textbf{151} & 153 \\
    21 & 6  &27 & 167 & 179 & \textbf{147} & 153 \\
    23 & 11 &84 & 335 & 407 & \textbf{323} & 357 \\
    31 & 5  &54 & 354 & \textbf{299} & 335 & 350 \\
    33 & 10 &85 & 413 & 440 & \textbf{404} & 434 \\
    35 & 12 &75 & 406 & 303 & 358 & \textbf{299} \\
    39 & 12 &97 & 502 & 425 & 472 & \textbf{391} \\
    45 & 12 &90 & 481 & 415 & 498 & \textbf{414} \\
    51 & 8  &115 & \textbf{641} & 755 & 676 & 739 \\
    63 & 6  &97  & 798 & \textbf{759} & 806 & 1031 \\
    65 & 12 &165 & 1092 & \textbf{901} & 1114 & 915 \\
    73 & 9  &144 & 1498 & 1567 & \textbf{1447} & 1526 \\
    85 & 8  &195 & 1601 & 1816 & \textbf{1589} & 1810 \\
    89 & 11 &336 & \textbf{2085} & 4326 & 2247 & 3973 \\
    91 & 12 &230 & 1668 & 1431 & 1596 & \textbf{1421} \\
    93 & 10 &223 & 1772 & 1939 & \textbf{1736} & 1788 \\
    105 &12 &234 & 1762 & 1481 & 1776 & \textbf{1333} \\
    117 & 12 &299 & 2304 & 2028 & 2366 & \textbf{1947} \\
    195 & 12 &496 & 4900 & 4230 & 4942 & \textbf{4166} \\
    273 & 12 &699 & 8064 & \textbf{7217} & 8082 & 7223 \\
    315 & 12 &752 & 8965 & \textbf{8032} & 9899  & 8099\\
    \hline
  \end{tabular}
  \label{tab:shortFFTs}
\end{table}

In the \textbf{third} step, we use the CFFTs with moderate lengths in
Table \ref{tab:shortFFTs} as sub-DFTs to construct our CCFTs.  With
\eqref{eq:additive} and \eqref{eq:multiplicative}, the computational
complexities of our CCFTs over GF$(2^l)$ ($4 \le l \le 12$) with
non-prime lengths can be calculated. The results are summarized in
Table \ref{tab:partI}, where the factorizations in parentheses are not
co-prime and the Cooley-Turkey algorithm is used in these cases. We
have tried all the decompositions with lengths smaller than 320, and
the decompositions with the smallest overall complexities are listed
in Table \ref{tab:partI}. Note that for each sub-DFT, the scheme with
the smallest additive complexity listed in Table \ref{tab:shortFFTs}
is used in the CCFT implementation to reduce the total additive
complexity. We also note that all DFT lengths in Table \ref{tab:partI}
are composite. The prime lengths are omitted because when $N$ is
prime, an $N$-point CCFT reduces to an $N$-point CFFT, which can be
found in Table \ref{tab:shortFFTs}.

\begin{table}[!htb]
  \centering
  \caption{The smallest complexity of our $N$-point CCFTs over
    GF$(2^{l})$  for composite $N$ and $N|2^l-1$ for $4\le l \le 12$ (we assume
    the sub-DFTs are shorter than 320).
  }
    \begin{tabular}{c|c|lccc}
      \hline
      $l$ & Length & Decomposition & mult. & add. & total\\
      \hline
      \multirow{1}{*}{4} & 15
      &  $1\times 15$ & 16 & 74 & 186 \\
      \hline
      \multirow{3}{*}{6}
      & 9 & $(3 \times 3)$ & 10 & 36 & 146\\
      \cline{2-6}
      & 21 & $3\times 7$ & 25 & 114 & 389\\
      \cline{2-6}
      & 63 & $(3 \times 3) \times 7$ & 124 & 468 & 1832\\
      \hline
      \multirow{3}{*}{8}
      & 51 &  $1 \times 51$ & 115 & 641 & 2366 \\
      \cline{2-6}
      & 85 &  $1 \times 85$ &195 & 1590 & 4515 \\
      \cline{2-6}
      & 255     & $3 \times 85$ & 670 & 5277 & 15327 \\
      \hline
      \multirow{1}{*}{9}
      & 511 & $7 \times 73$  & 1446 & 11881 & 36463 \\
      \hline
      \multirow{4}{*}{10}
      & 33 & $1\times 33$ & 85 & 404 & 2019 \\
      \cline{2-6}
      & 93 & $3 \times 31$ & 193 & 1083 & 4750 \\
      \cline{2-6}
      & 341&  $1 \times 341$ & 922  & 15184 & 32702\\
      \cline{2-6}
      & 1023 & $33 \times 31$           & 4417 & 22391 & 106314\\
      \hline
      \multirow{1}{*}{11}
      & 2047 & $23 \times 89$ & 15204 & 76702 & 395986\\
      \hline
      \multirow{15}{*}{12}
      & 35   & $5 \times 7$ & 65 & 232 & 1727\\
      \cline{2-6}
      & 39   & $1 \times 39$ & 97 & 391 & 2622 \\
      \cline{2-6}
      & 45   & $(3 \times 15)$ & 91 & 312 & 2405\\
      \cline{2-6}
      & 65   & $1 \times 65$ & 165 & 902 & 4697 \\
      \cline{2-6}
      & 91   & $1\times 93$ & 230 & 1421 & 6711 \\
      \cline{2-6}
      & 105  & $7 \times 15$    & 202 & 878 & 5524\\
      \cline{2-6}
      & 117  & $1 \times 117$ & 299 & 1947 & 8824 \\
      \cline{2-6}
      & 195  & $3 \times 65$     & 560 & 3093 & 15973\\
      \cline{2-6}
      & 273  & $ 3 \times 91$    & 781 & 4809 & 22772\\
      \cline{2-6}
      & 315  & $5 \times 63$     & 800 & 4803 & 23203\\
      \cline{2-6}
      & 455  & $7\times 65$      & 1545 & 7867 & 43402\\
      \cline{2-6}
      & 585  & $5 \times 117$    & 2080 & 11607 & 59447\\
      \cline{2-6}
      & 819  & $7 \times 117$        & 2795 & 16437 & 80722\\
      \cline{2-6}
      & 1365 & $7 \times 195$        & 4642 & 33842 &
      140608\\
      \cline{2-6}
      & 4095 & $65 \times 63$     & 16700 & 106098 & 490198\\
      \hline


    \end{tabular}
  \label{tab:partI}
\end{table}

Since some lengths of the DFTs have more than one
decomposition, it is possible that one decomposition scheme has a
smaller additive complexity but a larger multiplicative complexity
than another one.
Therefore, we need a metric to compare the overall complexities between
different decompositions.  In this paper, we follow our previous work
\cite{Chen2009c} and assume that the complexity of a multiplication
over GF$(2^l)$ is $2l-1$ times of that of an addition over the same
field, and the total complexity of a DFT is a weighted sum of the
additive and multiplicative complexities, i.e.,
$\mathrm{total}=(2l-1)\times \mathrm{mult} +\mathrm{add}$. This
assumption is based on both the software and hardware implementation
considerations \cite{Chen2009c}. Table \ref{tab:partI} lists the
decompositions with the smallest overall complexities.

Tables \ref{tab:partI} provide complexities of all $N$-point DFTs over
GF$(2^l)$ when $N | 2^l-1$ and $4 \le l \le 12$.  Note that the
decomposition corresponding to $1\times N$ is merely the $N$-point
CFFT over GF$(2^l)$. We have used the simplified CSE algorithm
described in Sec. \ref{sec:2047} to reduce the complexity of the
$2047$-point CFFTs over GF$(2^{11})$, and applied the CSE algorithm in
\cite{Chen2009c} to the other CFFTs. Thus, we have expanded the
results of \cite{Chen2009c}, where only the $(2^l-1)$-point CFFTs over
GF$(2^l)$ were given. We also observe that for some short lengths
(see, for example, $N=15$, $33$, or $65$), the $N$-point CFFTs lead to
the lowest complexity for the $N$-point CCFTs. For the DFTs with
lengths larger than 320, i.e., $511$-point CFFTs over GF$(2^9)$,
$341$-point CFFTs over GF$(2^{10})$, and 455-, $585$-, 819-, and
1365-point CFFTs over GF$(2^{12})$, the time complexity of the CSE
algorithm in \cite{Chen2009c} is still considerable. Thus, we cannot
minimize their complexities using schemes A, B, C, and D, and hence
they are not listed in Table \ref{tab:shortFFTs}.

Although the twiddle factors in the Cooley-Turkey algorithm
decomposition incur extra multiplicative complexity, Tables
\ref{tab:partI} show that the Cooley-Turkey algorithm decomposition
reduces the total complexity of our CCFTs in some cases (the
decompositions in parentheses). For example, while 9-point CFFT
requires 11 multiplications and 48 additions, $3\times 3$ CCFT based
on the Cooley-Turkey algorithm decomposition requires 10
multiplications and 36 additions. Despite the twiddle factors, the
CCFT based on the Cooley-Turkey algorithm decomposition have lower
multiplicative and additive complexities, because the Cooley-Turkey
algorithm decomposition allows us to take advantage of the low
complexity of the 3-point DFT.

\subsection{Complexity Comparison and Analysis}
\begin{table*}[!htb]
  \centering
  \caption{Comparison of the complexities our $N$-point CCFTs with
  FFTs available in the literature.}
\addtolength{\tabcolsep}{-3pt}
\scriptsize
  \begin{tabular}{|c|l|c|c|c|c|c|c|c|c|c|c|c|c|c|c|}
    \hline
    \multirow{3}{*}{$N$} & \multirow{3}{*}{Field} &
    \multicolumn{3}{c}{Wang and Zhu \cite{Wang1988}} &
    \multicolumn{3}{|c}{Trung et al. \cite{Truong2006}} &
    \multicolumn{5}{|c}{CFFT}
    & \multicolumn{3}{|c|}{CCFT} \\
    \cline{3-16}
    & & \multirow{2}{*}{mult.} & \multirow{2}{*}{add.} & \multirow{2}{*}{total} & \multirow{2}{*}{mult.} &
    \multirow{2}{*}{add.} & \multirow{2}{*}{total} &
    \multirow{2}{*}{mult.} & \multicolumn{2}{c|}{w/o CSE} &
    \multicolumn{2}{c|}{w/ CSE \cite{Chen2009c}} &
    \multirow{2}{*}{mult.} & \multirow{2}{*}{add.} &
    \multirow{2}{*}{total} \\
    \cline{10-13}
    &&&&&&&&&add. & total & add. & total &&&\\
    \hline
    15   & GF$(2^4)$ & 41    & 97    & 384  & --   & --    & --     & 16 & 201 &
    313 & 74    & \textbf{186}  & 20 & 78    & 218 \\
    63   & GF$(2^6)$ &801   & 801   & 9612   & --   & --    & --     & 97 & 2527 &
    3594 & 759   & \textbf{1826}   & 124   & 468   & 1832 \\
    255  & GF$(2^8)$ &1665  & 5377  & 30352  & 1135 & 3887  & 20902  & 586 &
    34783 & 43573 & 6736  & 15526  & 670   & 5277  & \textbf{15327}  \\
    511  & GF$(2^9)$ &13313 & 13313 & 239634 & 6516 & 17506 & 128278 & 1014 &
    141710 & 158948 & 23130 & 40368  & 1446  & 11881 & \textbf{36463}  \\
    1023 & GF$(2^{10})$ &32257 & 32257 & 645140 & 5915 & 30547 & 142932 & 2827 &
    536093 & 589806 & 75360 & 129073 & 4417  & 22391 & \textbf{106314} \\
    2047 & GF$(2^{11})$ &78601  & 78601 & 1689622     & --   & --    & --  & 7812 &
    2130248 & 2294300  & --  & --  & 15204 & 76702 & \textbf{395986} \\
    4095 & GF$(2^{12})$ &180225    & 180225    & 4325400     & --   & --    & -- &
    10832 & 8434414 & 8683550
    & --    & --     & 16700 & 106098 & \textbf{490198} \\
    \hline
  \end{tabular}
  \label{tab:comp}
\end{table*}

We compare the complexities of our CCFTs with those of previously
proposed FFTs in the literature in Table \ref{tab:comp}. For each
length, the lowest total complexity is in boldface font.  In Table
\ref{tab:comp}, our CCFTs achieve the lowest complexities for $N \geq
255$. Although the algorithm in \cite{Wang1988} is proved
asymptotically fast, the complexities of our CCFTs are only a fraction
of those in \cite{Wang1988}, and the advantage grows as the length
increases. Although the FFTs in \cite{Truong2006} are also based on
the prime-factor algorithm, our CCFTs achieve lower complexities for
two reasons. Since our CCFTs use CFFTs as the sub-DFTs, the
multiplicative complexities of our CCFTs are greatly reduced compared
with the FFTs in \cite{Truong2006}. For example, the multiplicative
complexity of our $511$-point CCFT is only one fourth of the
prime-factor algorithm in \cite{Truong2006}. Furthermore, using the
powerful CSE algorithm in \cite{Chen2009c}, the additive complexities
of our CCFTs are also greatly reduced.
Compared with the CFFTs, our CCFTs have a somewhat higher
multiplicative complexities, but this is more than made up by reduced
additive complexities of our CCFTs. The additive complexities of our
CCFTs are only a small fraction of those of CFFTs when directly
implemented. Compared with the CFFTs with reduced additive
complexities in \cite{Chen2009c}, our CCFTs still have much smaller
additive complexities due to their decomposition structure for $N \geq
63$. For example, the additive complexities of our CCFT is only about
half of that of the CFFT for $N=511$, and one third for $N=1023$. Due
to the significant reduction of the additive complexities, the total
complexities of our CCFTs with $N \geq 255$ are lower than those of
CFFTs. In comparison to CFFTs, the improvement by our CCFTs also grows
as the length increases.

For the DFTs whose lengths are prime, such as $31$-point DFT over
GF$(2^5)$, $127$-point DFT over GF$(2^7)$, and $8191$-point DFT over
GF$(2^{13})$, our CCFTs reduce to the CFFTs, and they have the same
computational complexities.

\section{Regular and Modular Structure of Our CCFTs}
\label{sec:arch}
We have shown that our CCFTs lead to lower complexities for moderate
to long lengths. Regardless of the length, our CCFTs also have
advantages in hardware implementations due to their regular and
modular structure.

\begin{figure}[!htb]
  \centering
  \includegraphics[width=8cm]{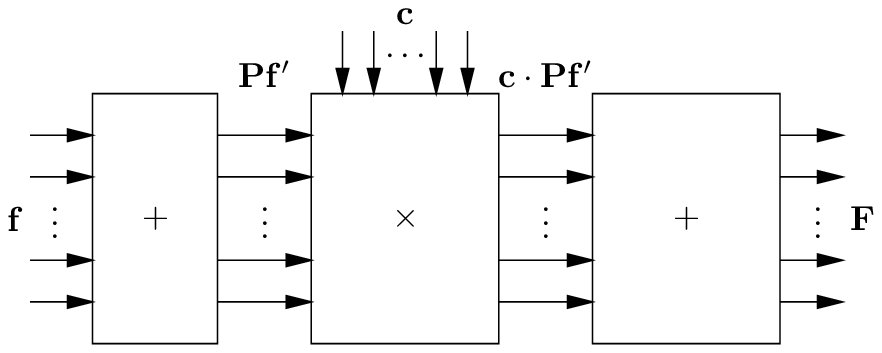}
  \caption{The structure of the CFFTs.}
  \label{fig:CFFT}
\end{figure}
The CFFT algorithm has a bilinear form, and therefore its circuitry
can be divided into three parts as shown in Fig.~\ref{fig:CFFT}. The
input vector $\mathbf{f}$ first goes through an pre-addition network,
which reorders $\mathbf{f}$ into $\mathbf{f}'$ and then computes
$\mathbf{Pf'}$. Then the resulting vector is sent to a multiplicative
network, in which the component-wise product of $\mathbf{c}$ and
$\mathbf{Pf}'$ is computed. The DFT result $\mathbf{F}$ is finally
computed in the post-addition network which corresponds to the linear
transform $\mathbf{AQ}$. While the structure of the CFFT looks simple,
the two additive networks are very complex for long DFTs. Although we
can reduce the additive complexity by the CSE algorithm, the resulted
additive networks still require a large number of
additions. Furthermore, the additions due to ${\A}$ or ${\A\Q}$ (the
second additive network in Fig.~\ref{fig:CFFT}) lack regularity, and
hence it is hard to use architectural techniques such as folding and
pipelining to achieve smaller area or high throughput.

\begin{figure}[!htb]
  \centering
  \includegraphics[width=5.5cm]{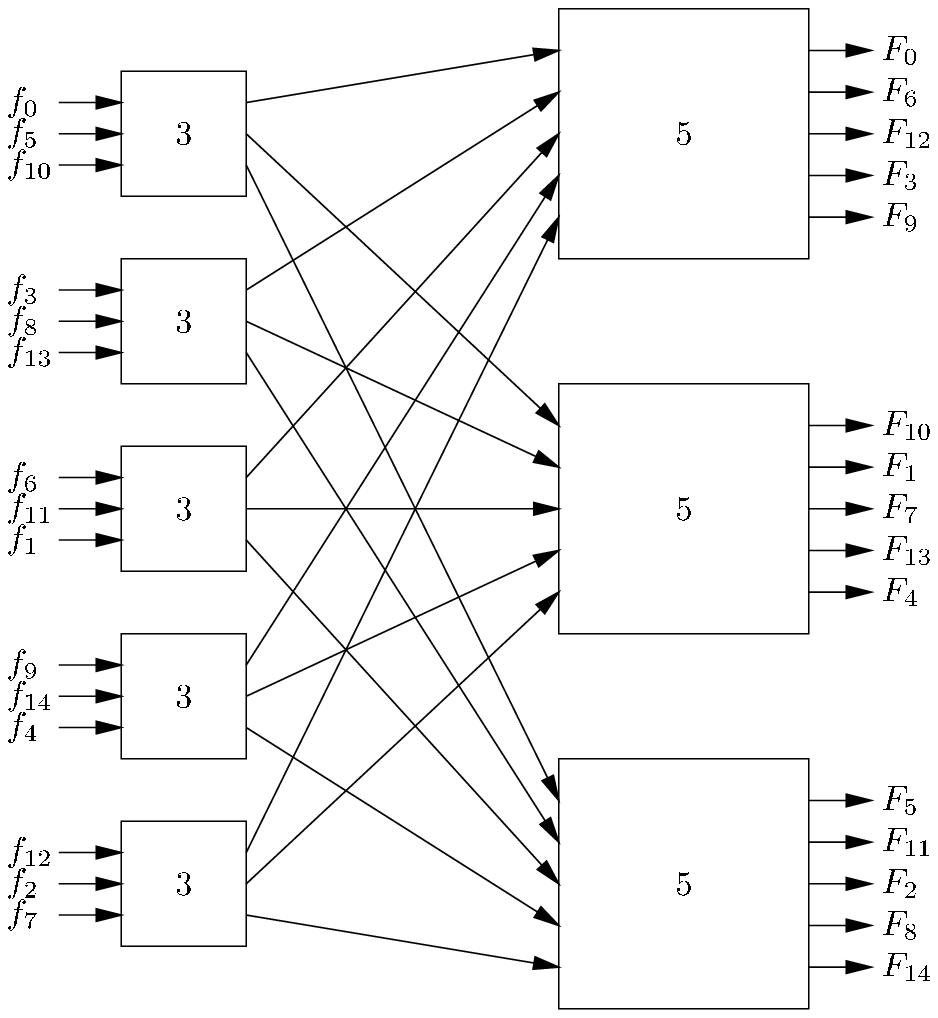}
  \caption{The regular and modular structure of our $15$-point CCFT based on a $3 \times 5$ decomposition. }
  \label{fig:CCFT3x5}
\end{figure}

In contrast, our CCFTs have regular and modular structure since they
are decomposed into shorter sub-DFTs. The sub-DFTs can be implemented
much easier than the long ones, and they can be reused in the CCFT
architecture. Fig.~\ref{fig:CCFT3x5} shows the regular and modular
structure of a $3 \times 5$ CCFT. Instead of designing the 15-point
CFFT directly, we only need to design a 3-point CFFT module and a
5-point CFFT module, and compute the 15-point CCFT by reusing these
modules according to the structure shown in Fig.~\ref{fig:CCFT3x5}.
It is much easier to apply architectural techniques such as folding
and pipelining to this regular and modular structure, leading to
efficient hardware implementations.

\section{Conclusion}
\label{sec:conclusion}
For any odd prime integer $p$, we reformulate $p$-point cyclic convolution
as a $(p-1)\times(p-1)$ Toeplitz matrix vector product, leading to
efficient cyclic convolution algorithms.  Based on this reformulation,
we have obtained efficient 11-point cyclic convolution algorithm and
derived the CFFTs over GF$(2^{11})$.  We have shown that our
composite cyclotomic Fourier transform algorithm leads to lower
complexities through decomposing long DFTs into shorter ones using
the prime-factor or Cooley-Turkey algorithms.
Our CCFTs over GF$(2^l)$ ($4\le l \le 12$), have lower complexities
than previously known FFTs over finite fields.  They also have a
regular and modular structure, which is desirable in hardware
implementations.

\appendices

\section{Short Toeplitz Matrix Vector Product over GF$(2^l)$}
\label{sec:TMVP}

An $n\times n$ TMVP over GF$(2^l)$ as
$$\begin{bmatrix}
  u_0\\
  u_1\\
  \vdots\\
  u_{n-1}
\end{bmatrix} =
\begin{bmatrix}
  r_{n-1} & r_n & \cdots  &r_{2n-2}\\
  r_{n-2} & r_{n-1} & \cdots & r_{2n-3} \\
  \vdots & \vdots & \ddots & \vdots\\
  r_{0} & r_1 & \cdots & r_{n-1}
\end{bmatrix}
\begin{bmatrix}
  v_0\\
  v_1\\
  \vdots\\
  v_{n-1}
\end{bmatrix}.
$$
can be computed with bilinear algorithm $\mathbf{E}^{(n)}(\mathbf{G}^{(n)}\mathbf{r}\cdot
\mathbf{H}^{(n)}\mathbf{v})$, where $\mathbf{r}=(r_0, r_1, \cdots,
r_{2n-2})^T$, $\mathbf{v}=(v_0,v_1,\cdots, v_{n-1})^T$, and
$\mathbf{E}^{(n)}$, $\mathbf{G}^{(n)}$ and $\mathbf{H}^{(n)}$ are all
binary matrices.

For $n=2$ (see, for example, \cite{Agarwal1974,Fan2007}),
$$
\mathbf{E}^{(2)}=
\begin{bmatrix}
  1 \, 0 \,  1\\
  0 \, 1 \,  1
\end{bmatrix}
,\;
\mathbf{G}^{(2)}=
\begin{bmatrix}
  0\, 1\, 1\\
  1\, 1\, 0 \\
  0\, 1\, 0
\end{bmatrix}, \;
\mathbf{H}^{(2)}=
\begin{bmatrix}
  0\, 1\\
  1\, 0\\
  1\, 1
\end{bmatrix}.
$$

For $n=3$  (see, for example, \cite{Fan2007}),
$$ (\mathbf{E}^{(3)})^T=
\begin{bmatrix}
  1\, 0\, 0 \\
  0\, 1\, 0 \\
  0\, 0\, 1 \\
  1\, 1\, 0 \\
  1\, 0\, 1 \\
  0\, 1\, 1 \\
\end{bmatrix},
\mathbf{G}^{(3)} =
\begin{bmatrix}
  0\, 0\, 1\, 1\, 1 \\
  0\, 1\, 1\, 1\, 0 \\
  1\, 1\, 1\, 0\, 0 \\
  0\, 0\, 0\, 1\, 0 \\
  0\, 0\, 1\, 0\, 0 \\
  0\, 1\, 0\, 0\, 0
\end{bmatrix},
\mathbf{H}^{(3)} =
\begin{bmatrix}
  0\, 0\, 1 \\
  0\, 1\, 0 \\
  1\, 0\, 0 \\
  0\, 1\, 1 \\
  1\, 0\, 1 \\
  1\, 1\, 0
\end{bmatrix}.
$$

For $n=5$,
$$
\mathbf{E}^{(5)} = \begin{bmatrix}
  1\, 1\, 1\, 1\, 1\, 0\, 0\, 0\, 0\\
  0\, 1\, 1\, 1\, 1\, 1\, 0\, 0\, 0\\
  0\, 0\, 1\, 1\, 1\, 1\, 1\, 0\, 0\\
  0\, 0\, 0\, 1\, 1\, 1\, 1\, 1\, 0\\
  0\, 0\, 0\, 0\, 1\, 1\, 1\, 1\, 1\\
  0\, 1\, 0\, 0\, 1\, 0\, 0\, 0\, 0\\
  0\, 0\, 1\, 0\, 0\, 0\, 0\, 0\, 0\\
  0\, 0\, 0\, 1\, 1\, 0\, 0\, 0\, 0\\
  0\, 0\, 0\, 1\, 0\, 0\, 0\, 0\, 0\\
  0\, 0\, 0\, 0\, 1\, 1\, 0\, 0\, 0\\
  0\, 0\, 0\, 0\, 0\, 1\, 0\, 0\, 0\\
  0\, 0\, 0\, 0\, 0\, 0\, 1\, 0\, 0\\
  0\, 0\, 0\, 0\, 1\, 0\, 0\, 1\, 0\\
  0\, 0\, 0\, 0\, 1\, 0\, 0\, 0\, 0
\end{bmatrix}, \quad
\mathbf{G}^{(5)} = \begin{bmatrix}
  1\, 0\, 0\, 0\, 0\\
  0\, 1\, 0\, 0\, 0\\
  0\, 0\, 1\, 0\, 0\\
  0\, 0\, 0\, 1\, 0\\
  0\, 0\, 0\, 0\, 1\\
  1\, 1\, 0\, 0\, 0\\
  1\, 0\, 1\, 0\, 0\\
  1\, 0\, 0\, 1\, 0\\
  0\, 1\, 1\, 0\, 0\\
  0\, 1\, 0\, 0\, 1\\
  0\, 0\, 1\, 1\, 0\\
  0\, 0\, 1\, 0\, 1\\
  0\, 0\, 0\, 1\, 1\\
  1\, 1\, 0\, 1\, 1
\end{bmatrix},
$$
$$
\mathbf{H}^{(5)} = \begin{bmatrix}
  0\, 0\, 0\, 0\, 1\, 0\, 0\, 0\, 0\, 1\, 0\, 1\, 1\, 1\\
  0\, 0\, 0\, 1\, 0\, 0\, 0\, 1\, 0\, 0\, 1\, 0\, 1\, 1\\
  0\, 0\, 1\, 0\, 0\, 0\, 1\, 0\, 1\, 0\, 1\, 1\, 0\, 0\\
  0\, 1\, 0\, 0\, 0\, 1\, 0\, 0\, 1\, 1\, 0\, 0\, 0\, 1\\
  1\, 0\, 0\, 0\, 0\, 1\, 1\, 1\, 0\, 0\, 0\, 0\, 0\, 1
\end{bmatrix}.
$$

\section{4-, 8-, and 11-point Cyclic Convolution Algorithms over
  GF$(2^l)$}
\label{sec:convs}

For 4-point cyclic convolutions, \cite{Afanasyev1987}
$$
\mathbf{Q}^{(4)}=\left[
  \begin{array}{ccccccccc}
    1&1&1&1&0&0&0&0&0\\
    1&1&0&0&1&1&1&1&0\\
    1&1&1&1&1&1&0&0&0\\
    1&1&0&0&0&0&1&1&1\\
  \end{array}
\right],
$$
$$\mathbf{R}^{(4)}=\left[
  \begin{array}{cccc}
    1&1&0&0\\
    1&1&1&1\\
    1&0&1&0\\
    0&1&0&0\\
    1&1&1&1\\
    0&1&0&1\\
    0&1&0&1\\
    0&0&1&0\\
    1&1&1&1
  \end{array}
\right], \mathbf{P}^{(4)}=\left[
  \begin{array}{cccc}
    1&0&1&0\\
    0&1&0&0\\
    0&1&1&0\\
    1&1&1&1\\
    1&0&1&0\\
    1&1&1&1\\
    0&0&1&1\\
    1&1&1&1\\
    1&1&1&1\\
  \end{array}
\right].
$$

For 8-point cyclic convolution \cite{Churkov2004},
$$\mathbf{Q}^{(8)}=
  \begin{bmatrix}
    1\, 1\, 0\, 0\, 0\, 0\, 1\, 1\, 0\, 1\, 1\, 0\, 0\, 0\, 0\, 0\, 0\, 0\, 0\, 0\, 0\, 0\, 0\, 0\, 1\, 1\, 0\\
    1\, 0\, 1\, 0\, 0\, 0\, 1\, 0\, 1\, 1\, 0\, 1\, 0\, 0\, 0\, 0\, 0\, 0\, 0\, 0\, 0\, 0\, 0\, 0\, 1\, 1\, 1\\
    1\, 1\, 0\, 0\, 0\, 0\, 1\, 1\, 0\, 0\, 0\, 0\, 1\, 1\, 0\, 0\, 0\, 0\, 0\, 0\, 0\, 1\, 1\, 0\, 0\, 0\, 0\\
    1\, 0\, 1\, 0\, 0\, 0\, 1\, 0\, 1\, 0\, 0\, 0\, 1\, 0\, 1\, 0\, 0\, 0\, 0\, 0\, 0\, 1\, 1\, 1\, 0\, 0\, 0\\
    1\, 1\, 0\, 1\, 1\, 0\, 1\, 1\, 0\, 1\, 1\, 0\, 0\, 0\, 0\, 0\, 0\, 0\, 1\, 1\, 0\, 0\, 0\, 0\, 0\, 0\, 0\\
    1\, 0\, 1\, 1\, 0\, 1\, 1\, 0\, 1\, 1\, 0\, 1\, 0\, 0\, 0\, 0\, 0\, 0\, 1\, 1\, 1\, 0\, 0\, 0\, 0\, 0\, 0\\
    1\, 1\, 0\, 1\, 1\, 0\, 1\, 1\, 0\, 0\, 0\, 0\, 1\, 1\, 0\, 1\, 1\, 0\, 0\, 0\, 0\, 0\, 0\, 0\, 0\, 0\, 0\\
    1\, 0\, 1\, 1\, 0\, 1\, 1\, 0\, 1\, 0\, 0\, 0\, 1\, 0\, 1\, 1\, 1\, 1\, 0\, 0\, 0\, 0\, 0\, 0\, 0\, 0\, 0
  \end{bmatrix},
$$
$$
(\mathbf{R}^{(8)})^T=
  \begin{bmatrix}
    1\, 1\, 1\, 1\, 1\, 1\, 1\, 1\, 1\, 1\, 1\, 1\, 1\, 1\, 1\, 1\, 1\, 1\, 1\, 1\, 1\, 1\, 1\, 1\, 1\, 1\, 1\\
    0\, 0\, 1\, 0\, 0\, 1\, 0\, 0\, 1\, 0\, 0\, 1\, 0\, 1\, 1\, 0\, 1\, 1\, 0\, 1\, 1\, 0\, 1\, 1\, 0\, 1\, 1\\
    0\, 0\, 0\, 0\, 0\, 0\, 0\, 0\, 0\, 0\, 0\, 0\, 1\, 1\, 1\, 1\, 1\, 1\, 1\, 1\, 1\, 1\, 1\, 1\, 1\, 1\, 1\\
    0\, 0\, 0\, 0\, 1\, 0\, 0\, 1\, 0\, 0\, 0\, 0\, 0\, 0\, 1\, 0\, 1\, 1\, 0\, 1\, 1\, 0\, 1\, 1\, 0\, 1\, 1\\
    0\, 0\, 0\, 1\, 1\, 1\, 1\, 1\, 1\, 0\, 0\, 0\, 0\, 0\, 0\, 1\, 1\, 1\, 1\, 1\, 1\, 1\, 1\, 1\, 1\, 1\, 1\\
    0\, 0\, 0\, 0\, 0\, 1\, 0\, 0\, 1\, 0\, 1\, 0\, 0\, 0\, 0\, 0\, 1\, 1\, 0\, 1\, 1\, 0\, 1\, 1\, 0\, 1\, 1\\
    0\, 0\, 0\, 0\, 0\, 0\, 0\, 0\, 0\, 1\, 1\, 1\, 0\, 0\, 0\, 1\, 1\, 1\, 1\, 1\, 1\, 1\, 1\, 1\, 1\, 1\, 1\\
    0\, 1\, 0\, 0\, 1\, 0\, 0\, 1\, 0\, 0\, 1\, 1\, 0\, 1\, 0\, 0\, 1\, 1\, 0\, 1\, 1\, 0\, 1\, 1\, 0\, 1\, 1\\
  \end{bmatrix},
$$
and
$$(\mathbf{P}^{(8)})^T=
  \begin{bmatrix}
    1\, 0\, 1\, 1\, 0\, 1\, 0\, 0\, 0\, 0\, 0\, 0\, 1\, 0\, 1\, 1\, 0\, 1\, 0\, 0\, 0\, 0\, 0\, 0\, 0\, 0\, 0\\
    1\, 1\, 0\, 1\, 1\, 0\, 0\, 0\, 0\, 0\, 0\, 0\, 1\, 1\, 0\, 1\, 1\, 1\, 0\, 0\, 0\, 0\, 0\, 0\, 0\, 0\, 0\\
    1\, 0\, 1\, 1\, 0\, 1\, 0\, 0\, 0\, 1\, 0\, 1\, 0\, 0\, 0\, 0\, 0\, 0\, 1\, 0\, 1\, 0\, 0\, 0\, 0\, 0\, 0\\
    1\, 1\, 0\, 1\, 1\, 0\, 0\, 0\, 0\, 1\, 1\, 0\, 0\, 0\, 0\, 0\, 0\, 0\, 1\, 1\, 1\, 0\, 0\, 0\, 0\, 0\, 0\\
    1\, 0\, 1\, 1\, 0\, 1\, 1\, 0\, 1\, 0\, 0\, 0\, 1\, 0\, 1\, 0\, 0\, 0\, 0\, 0\, 0\, 1\, 0\, 1\, 0\, 0\, 0\\
    1\, 1\, 0\, 1\, 1\, 0\, 1\, 1\, 0\, 0\, 0\, 0\, 1\, 1\, 0\, 0\, 0\, 0\, 0\, 0\, 0\, 1\, 1\, 1\, 0\, 0\, 0\\
    1\, 0\, 1\, 1\, 0\, 1\, 1\, 0\, 1\, 1\, 0\, 1\, 0\, 0\, 0\, 0\, 0\, 0\, 0\, 0\, 0\, 0\, 0\, 0\, 1\, 0\, 1\\
    1\, 1\, 0\, 1\, 1\, 0\, 1\, 1\, 0\, 1\, 1\, 0\, 0\, 0\, 0\, 0\, 0\, 0\, 0\, 0\, 0\, 0\, 0\, 0\, 1\, 1\, 1\\
  \end{bmatrix}.
$$

For our new 11-point cyclic convolutions, $\mathbf{Q}^{(11)}$ is given by
$$
  \begin{bmatrix}
    1 0 0 0 0 0 0 0 0 0 0 0 0 0 0 1 1 1 1 1 0 0 0 0 0 0 0 0 0 1 1 1 1 1 0 0 0 0 0 0 0 0 0\\
    1 0 0 0 0 1 0 0 0 0 1 0 1 1 1 0 0 0 0 1 0 0 0 0 1 0 1 1 1 0 0 0 0 0 0 0 0 0 0 0 0 0 0\\
    1 0 0 0 1 0 0 0 1 0 0 1 0 1 1 0 0 0 1 0 0 0 1 0 0 1 0 1 1 0 0 0 0 0 0 0 0 0 0 0 0 0 0\\
    1 0 0 1 0 0 0 1 0 1 0 1 1 0 0 0 0 1 0 0 0 1 0 1 0 1 1 0 0 0 0 0 0 0 0 0 0 0 0 0 0 0 0\\
    1 0 1 0 0 0 1 0 0 1 1 0 0 0 1 0 1 0 0 0 1 0 0 1 1 0 0 0 1 0 0 0 0 0 0 0 0 0 0 0 0 0 0\\
    1 1 0 0 0 0 1 1 1 0 0 0 0 0 1 1 0 0 0 0 1 1 1 0 0 0 0 0 1 0 0 0 0 0 0 0 0 0 0 0 0 0 0\\
    1 0 0 0 0 1 0 0 0 0 1 0 1 1 1 0 0 0 0 0 0 0 0 0 0 0 0 0 0 0 0 0 0 1 0 0 0 0 1 0 1 1 1\\
    1 0 0 0 1 0 0 0 1 0 0 1 0 1 1 0 0 0 0 0 0 0 0 0 0 0 0 0 0 0 0 0 1 0 0 0 1 0 0 1 0 1 1\\
    1 0 0 1 0 0 0 1 0 1 0 1 1 0 0 0 0 0 0 0 0 0 0 0 0 0 0 0 0 0 0 1 0 0 0 1 0 1 0 1 1 0 0\\
    1 0 1 0 0 0 1 0 0 1 1 0 0 0 1 0 0 0 0 0 0 0 0 0 0 0 0 0 0 0 1 0 0 0 1 0 0 1 1 0 0 0 1\\
    1 1 0 0 0 0 1 1 1 0 0 0 0 0 1 0 0 0 0 0 0 0 0 0 0 0 0 0 0 1 0 0 0 0 1 1 1 0 0 0 0 0 1\\
  \end{bmatrix},
$$
the transpose of $\mathbf{R}^{(11)}$ is given by
$$
  \begin{bmatrix}
    1 1 0 0 0 0 0 1 0 1 1 0 1 0 1 1 1 1 1 1 0 0 0 0 1 1 0 0 0 0 1 1 1 1 0 0 0 0 1 1 0 0 0\\
    1 0 0 0 0 0 1 1 1 1 1 1 1 1 0 1 1 1 1 1 1 0 1 0 1 0 0 1 1 1 1 1 1 1 1 0 1 0 1 0 0 1 1\\
    1 0 0 0 0 1 0 1 1 0 0 1 1 0 1 1 1 1 1 0 0 0 1 1 0 0 0 0 0 1 1 1 1 1 0 0 1 1 0 0 0 0 0\\
    1 0 0 0 1 1 0 0 0 1 0 1 1 0 1 1 1 1 0 1 0 1 0 0 0 0 0 0 0 1 1 1 1 1 0 1 0 0 0 0 0 1 0\\
    1 0 0 1 1 1 1 1 0 1 0 1 1 0 1 1 1 0 1 1 1 0 0 0 0 0 0 1 0 1 1 1 1 1 1 0 0 0 0 0 1 0 0\\
    1 0 1 1 1 1 0 1 0 1 0 1 1 0 1 1 0 1 1 1 0 0 0 0 0 0 1 0 0 1 1 1 1 1 0 0 0 0 1 1 0 0 0\\
    1 1 1 1 1 1 0 1 0 1 0 1 1 0 1 0 1 1 1 1 0 0 0 0 1 1 0 0 0 1 1 1 1 1 1 0 1 0 1 0 0 1 1\\
    1 1 1 1 1 1 0 1 0 1 0 1 1 0 1 1 1 1 1 1 1 0 1 0 1 0 0 1 1 1 1 1 1 0 0 0 1 1 0 0 0 0 0\\
    1 1 1 1 1 0 0 1 0 1 0 1 1 0 1 1 1 1 1 1 0 0 1 1 0 0 0 0 0 1 1 1 0 1 0 1 0 0 0 0 0 0 0\\
    1 1 1 1 0 0 0 1 0 1 0 1 1 1 1 1 1 1 1 1 0 1 0 0 0 0 0 1 0 1 1 0 1 1 1 0 0 0 0 0 0 1 0\\
    1 1 1 0 0 0 0 1 0 1 0 1 0 0 1 1 1 1 1 1 1 0 0 0 0 0 1 0 0 1 0 1 1 1 0 0 0 0 0 0 1 0 0\\
  \end{bmatrix},
$$
and the transpose of $\mathbf{P}^{(11)}$ is given by
$$
  \begin{bmatrix}
    1 1 0 0 0 0 1 1 1 0 0 0 0 0 1 0 0 0 0 0 0 0 0 0 0 0 0 0 0 1 0 0 0 0 1 1 1 0 0 0 0 0 1\\
    1 0 1 0 0 0 1 0 0 1 1 0 0 0 1 0 0 0 0 0 0 0 0 0 0 0 0 0 0 0 1 0 0 0 1 0 0 1 1 0 0 0 1\\
    1 0 0 1 0 0 0 1 0 1 0 1 1 0 0 0 0 0 0 0 0 0 0 0 0 0 0 0 0 0 0 1 0 0 0 1 0 1 0 1 1 0 0\\
    1 0 0 0 1 0 0 0 1 0 0 1 0 1 1 0 0 0 0 0 0 0 0 0 0 0 0 0 0 0 0 0 1 0 0 0 1 0 0 1 0 1 1\\
    1 0 0 0 0 1 0 0 0 0 1 0 1 1 1 0 0 0 0 0 0 0 0 0 0 0 0 0 0 0 0 0 0 1 0 0 0 0 1 0 1 1 1\\
    1 1 0 0 0 0 1 1 1 0 0 0 0 0 1 1 0 0 0 0 1 1 1 0 0 0 0 0 1 0 0 0 0 0 0 0 0 0 0 0 0 0 0\\
    1 0 1 0 0 0 1 0 0 1 1 0 0 0 1 0 1 0 0 0 1 0 0 1 1 0 0 0 1 0 0 0 0 0 0 0 0 0 0 0 0 0 0\\
    1 0 0 1 0 0 0 1 0 1 0 1 1 0 0 0 0 1 0 0 0 1 0 1 0 1 1 0 0 0 0 0 0 0 0 0 0 0 0 0 0 0 0\\
    1 0 0 0 1 0 0 0 1 0 0 1 0 1 1 0 0 0 1 0 0 0 1 0 0 1 0 1 1 0 0 0 0 0 0 0 0 0 0 0 0 0 0\\
    1 0 0 0 0 1 0 0 0 0 1 0 1 1 1 0 0 0 0 1 0 0 0 0 1 0 1 1 1 0 0 0 0 0 0 0 0 0 0 0 0 0 0\\
    1 0 0 0 0 0 0 0 0 0 0 0 0 0 0 1 1 1 1 1 0 0 0 0 0 0 0 0 0 1 1 1 1 1 0 0 0 0 0 0 0 0 0\\
  \end{bmatrix}.
$$

\bibliographystyle{IEEEtran} \bibliography{IEEEabrv,FFT}

\end{document}